\documentclass[preprint,prd,aps,showpacs,nofootinbib]{revtex4}
\usepackage{inputenc}
\usepackage{epsfig,float}
\usepackage{amsfonts}
\usepackage[hypertex]{hyperref}
\usepackage{amsmath}
\input{epsf}
\newcommand{\be}{\begin{eqnarray}}
\newcommand{\ee}{\end{eqnarray}}
\newcommand{\ba}{\begin{array}}
\newcommand{\ea}{\end{array}}

\newcommand{\ks}[1]{{\rlap/ #1}}
\restylefloat{figure}
\restylefloat{table}

\begin{document}

\title{A spectral representation for  baryon to meson
%and baryon to photon
transition distribution
amplitudes}

\author{B.~Pire$^1$,  K.~Semenov-Tian-Shansky$^{1,2}$, L.~Szymanowski$^{3}$ }
\affiliation{$^1$ CPhT, \'{E}cole Polytechnique, CNRS,  91128, Palaiseau, France  \\
$^2$ LPT,   Universit\'{e} d'Orsay, CNRS, 91404 Orsay, France \\
%$^3$ St. Petersburg State University, St.Petersburg,  \\ 198504, Petrodvoretz,   Russia \\
$^3$ Soltan Institute for Nuclear Studies, Warsaw,  Poland.
}
%\email[]{Kirill.Semenov@cpht.polytechnique.fr}

\preprint{CPHT-RR051.0610, LPT-ORSAY 10-47}

\begin{abstract}

We construct a spectral representation for the
baron to meson
transition distribution amplitudes
(TDAs), {\it i.e.}
matrix elements involving
three quark correlators which arise
%{\it e.g.}
in the
description of baryon to meson
%and baryon to photon
transitions within the factorization approach
to hard exclusive reactions. We generalize for
these quantities the notion of double distributions introduced in the context of
generalized parton distributions. We propose the generalization
of A.~Radyushkin's factorized Ansatz for the case  of baryon to meson
%and baryon to photon
TDAs.
 Our construction opens the way to
modeling of baryon to meson
%and baryon to photon
TDAs
in their complete domain of
definition and quantitative
estimates of cross-sections for various hard exclusive reactions.

\end{abstract}

\maketitle

\renewcommand{\thesection}{\arabic{section}}

\renewcommand{\thesubsection}{\arabic{section}.\arabic{subsection}}
\renewcommand{\thefigure}{\arabic{figure}}

\section{Introduction}

The concept of generalized parton distributions (GPDs)
\cite{pioneers0,pioneers1,pioneers2,pioneers3},
which in the simplest (leading twist) case are non-diagonal matrix elements of quark-antiquark or gluon-gluon
non local operators on the light cone, has recently been extended
\cite{Frankfurt:1999fp,Frankfurt:2002kz}
to baryon to meson
(and baryon to photon)
transition distribution amplitudes (TDAs), non diagonal matrix elements
of three quark operators between two hadronic states of different baryon number
(or between a baryon state and a photon).
Nucleon to meson TDAs are conceptually much related to meson-nucleon generalized distribution amplitudes
\cite{BLP1,BLP2} since they involve the same non-local operators
\cite{Radyushkin:1977gp,Efremov:1978rn,Lepage:1980,Chernyak:1983ej}.
These objects are useful for the description of exclusive processes characterized by
a baryonic exchange such as
%backward deeply virtual Compton scattering
%\cite{PS phys rev},
backward electroproduction of mesons
\cite{Pire:2005ax,Pire:2005mt,Lansberg:2007ec}
or proton-antiproton hard exclusive annihilation processes
\cite{LPS}.
Nucleon to meson TDAs are also considered to be a useful tool to quantify  the pion cloud in %the
baryons
\cite{Strikman:2010pu}.

Up to now  TDAs between the states of unequal baryon  number lacked any suitable
phenomenological parametrization in the whole domain of their definition, as for
example in the framework of the quark model developed in
\cite{Pasquini:2009ki}.
The complete parametrization should properly take into account the
fundamental requirement of Lorentz covariance which is manifest as the polynomiality
property of the Mellin moments in the relevant light-cone momentum fraction on the complete
domain of their definition.
For the case of the  GPDs an elegant way to fulfill this requirement
consists in employing the spectral representation. The corresponding spectral properties
were established with the help of the alpha-representation techniques
\cite{Radyushkin:1983ea, Radyushkin:1983wh}.
Radyushkin's factorized Ansatz
based on the double distribution representation for GPDs
\cite{RDDA1,RDDA2,RDDA3,RDDA4}
became the basis for various
successful phenomenological GPD models
(see \cite{GPV,Diehl,BelRad,Boffi:2007yc,GVV}).

In this paper we address the problem of construction
of a spectral representation of baryon to meson
%(and baryon to photon)
transition distribution amplitudes.
We introduce the notion of quadruple distributions and generalize  Radyushkin's
factorized Ansatz for this issue.
This allows the  modeling of baryon to meson
%and baryon to photon
TDAs
in the complete domain of their definition and quantitative
rate estimates in various hard exclusive reactions.

Similarly as the nucleon to meson TDAs factorize in backward meson
electroproduction, nucleon to photon TDAs may factorize in backward
virtual Compton scattering
\cite{PS phys rev}. The main part of the analysis performed in our paper
can be directly applied to the nucleon to photon TDAs. But the modelling
of the quadruple distribution has to account for the anomalous nature
of a photon. The studies of the anomalous photon structure functions
\cite{Witten}
and of the photon GPDs
\cite{Friot}
show that taking it into account is
a nontrivial task which deserves
separate studies.

\section{Basic definitions and kinematics}
%\subsection{}

\begin{figure}[h]
 \begin{center}
 \epsfig{figure=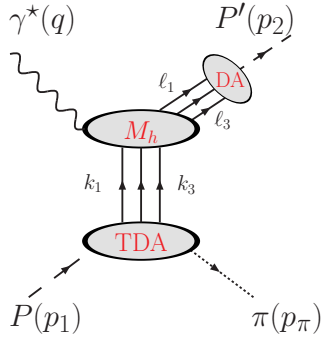 , height=4.5cm}
  \caption{The factorization of the process $\gamma^*+P \rightarrow P'+\pi$.
The lower blob is the pion-nucleon transition distribution amplitude,
$M_h$ denotes the hard subprocess amplitude, DA is the nucleon distribution amplitude. }
\label{Kinematics}
\end{center}
\end{figure}

Nucleon to meson transition distribution amplitudes
also called in the literature as skewed DAs
\cite{Frankfurt:1999fp} and superskewed parton distributions
\cite{Frankfurt:2002kz}
which extend the concept of usual generalized parton distributions
arise {\it e.g.} in the description of   meson electroproduction on the nucleon target
\cite{Pire:2005ax, Pire:2005mt, Lansberg:2007ec}. For definiteness below we consider the
case of nucleon to pion transition distribution amplitudes ($\pi N$ TDAs for brevity)
although our analysis is general enough to be applied to other baryon-meson and
also baryon to photon TDAs.
$\pi N$ TDAs arise
in the description of backward pion electroproduction
\be
\gamma^*(q)\, + \, N(p_1) \, \rightarrow N'(p_2)\, +\, \pi (p_\pi)\,,
\label{Backward_pi_prod}
\ee
in the generalized Bjorken regime
($-q^2$--large;
$q^2/(2p_1 \cdot q)$ kept fixed; $-q^2 \gg -u$).
The factorization theorem
was argued for the process
(\ref{Backward_pi_prod})
in \cite{Frankfurt:1999fp,Frankfurt:2002kz}
 (see Fig.~\ref{Kinematics}).
The appropriate kinematics is described as follows
\cite{Lansberg:2007ec}:
\be
&&
P=\frac{1}{2}(p_1+p_\pi)\,; \ \ \ \Delta=p_\pi-p_1\,; \nonumber \\ &&
u=\Delta^2\,; \ \ \ \xi=-\frac{\Delta \cdot n}{2 P \cdot n} \,,
\ee
where $u$ denotes the transfer momentum squared between the meson and the nucleon target
and $\xi$
is the skewness parameter.
$n$
and
$p$
are the usual light-cone vectors occurring in the Sudakov decomposition of momenta
($n^2=p^2=0$, $n \cdot p=1$).
The light-cone decomposition of the particular vector
$v^\mu$
is given by
$v^\mu=v^+ p^\mu+ v^- n^\mu+v_T^\mu$.
%$n$ and $\tilde{p}$ are the light-cone direction ($P \cdot n = P^+$ and )

The definition of $\pi N$
TDAs can be symbolically written as \cite{Frankfurt:1999fp,Frankfurt:2002kz}:
\be
&&
\int  \left[ \prod_{i=1}^3 \frac{dz^-_i}{2 \pi} \right]
e^{ix_1(P \cdot z_1)+ix_2(P \cdot z_2)+ix_3(P \cdot z_3)} \nonumber \\ &&
\times
\left.
\langle \pi(P+\Delta/2) |
\epsilon_{abc}
\psi_{j_1}^a(z_1)
\psi_{j_2}^b(z_2)
\psi_{j_3}^c(z_3)
|N(P-\Delta/2) \rangle
\right|_{z_i^+=z_i^\bot=0}
 \nonumber \\ &&
\sim \delta(2 \xi-x_1-x_2-x_3) H_{j_1 \, j_2 \, j_3} (x_1,\,x_2,\,x_3,\, \xi, \, u)\,.
 \label{TDA_definition}
\ee
Here $j_{1,2,3}$ stand for spin-flavor indices and $a$, $b$, $c$ are color indices.
The decomposition of the Fourier transform
(\ref{TDA_definition})
of the matrix element  of the three-local light-cone  quark operator
involves a  set of independent spin-flavor structures multiplied by corresponding invariant functions:
$\pi N$
TDAs.

It is worth to mention that in order to preserve gauge invariance one has to insert
the path-ordered gluonic exponentials
$[z_i;z_0]$
along the straight line connecting an arbitrary initial point
$z_0 n$
and a final one
$z_i n$:
\be
   \langle \pi|\, \epsilon^{abc} \psi_{j_1}^{a'} (z_{1} )\,
[z_1;z_0]_{a',a}\,\psi_{j_2}^{b'} (z_{2} )\, [z_2;z_0]_{b'b}\,
   \psi_{j_3}^{c'} (z_{3}  ) [z_3;z_0]_{c'c} \,|N \rangle.
\ee
Throughout this paper we adopt the light-cone
gauge $A^+=0$, so that the gauge link is equal to unity. Thus we do not show it explicitly
in the definition (\ref{TDA_definition}).

For the case of proton to $\pi^0$ transition
the decomposition of
(\ref{TDA_definition})
over the independent spinor structures at the leading twist
involves $8$ independent terms.
It reads%
\footnote{We make use of the notation
${\cal F}(\cdot)= (P \cdot n)^3 \int \left[ \prod_{i=1}^3  \frac{dz_i}{2 \pi} \right]
e^{ix_1(P \cdot z_1)+ix_2(P \cdot z_2)+ix_3(P \cdot z_3)}
(\cdot)$
}
\cite{Lansberg:2007ec}:
%%%%%%%%%%%%%%%%%%%%%%%%%%%%%%%%%%%%%%%%%%%%%%%%%%%%%%%%%%%%%%%%
\be
 && 4 {\cal F}\Big(\langle     \pi^0(p_\pi)|\, \epsilon_{abc}u^{a}_{\alpha}(z_1 n)
u^{b}_{\beta}(z_2 n)d^{c}_{\gamma}(z_3 n)
\,|P(p_1,s_1) \rangle \Big) \nonumber
 = \delta(2 \xi-x_1-x_2-x_3)
\\ \nonumber
&&
\times i\frac{f_N}{f_\pi} \Big[
V^{p\pi^0}_{1} (\ks p C)_{\alpha\beta}(N^+)_{\gamma}+A^{p\pi^0}_{1} (\ks p\gamma^5 C)_{\alpha\beta}(\gamma^5 N^+)_{\gamma} +
T^{p\pi^0}_{1} (\sigma_{p\mu} C)_{\alpha\beta}(\gamma^\mu N^+)_{\gamma}
\nonumber \\
&& + M^{-1}V^{p\pi^0}_{2}
 (\ks p C)_{\alpha\beta}(\ks \Delta\!_T N^+)_{\gamma} +M^{-1}
A^{p\pi^0}_{2}(\ks p \gamma^5 C)_{\alpha\beta}(\gamma^5\ks \Delta\!_T N^+)_{\gamma}
+ M^{-1}T^{p\pi^0}_{2} ( \sigma_{p\Delta_T} C)_{\alpha\beta}(N^+)_{\gamma}
\nonumber \\
&&+  M^{-1}T^{p\pi^0}_{3} ( \sigma_{p\mu} C)_{\alpha\beta}(\sigma^{\mu\Delta_T}
 N^+)_{\gamma} + M^{-2}T^{p\pi^0}_{4} (\sigma_{p \Delta_T} C)_{\alpha\beta}
(\ks \Delta\!_T N^+)_{\gamma}\;\Big]. %\nonumber
%\end{eqnarray}
\label{pi0p_TDA_decomp}
\ee
Here $\ks{p}$ is the usual Dirac slash notation
($\ks{p}=p_\mu \gamma^\mu$),
$\sigma^{\mu\nu}= \frac{1}{2}[\gamma^\mu, \gamma^\nu]$
with
$\sigma^{p \mu} = p_\nu \sigma^{\nu\mu}$,
 $C$
is the charge conjugation matrix
and $N^+$ is the large component of the nucleon spinor
($N=(\ks n \ks p + \ks p \ks n) N = N^-+N^+$
with $N^+\sim \sqrt{p_1^+}$ and $N^-\sim \sqrt{1/p_1^+}$).
$M$ stands for the nucleon mass,
$f_\pi$ is the pion decay constant ($f_\pi = 131$ MeV) and $f_N$ is a constant
with the dimension of energy squared. All the
$8$ $p \rightarrow \pi^0$ TDAs $V_i$, $A_i$ and $T_i$ are
dimensionless.
%Only first three
%terms in (\ref{pi0p_TDA_decomp})  survive the limit $\Delta_T \to 0$.

%%%%%%%%%%%%%%%%%%%%%%%%%%%%%%%%%%%%%%%%%%%%%%%%%%%%%%%%%%%%%%%%%%

In this paper we %put aside the problem of proper Lorentz and flavor decomposition
%of the r.h.s. of
%(\ref{TDA_definition}).
%We rather
concentrate on the dependence of the invariant functions
$V_i$, $A_i$, $T_i$
multiplying the independent spin-flavor structures in
(\ref{pi0p_TDA_decomp})
on the longitudinal momentum fractions
$x_1$, $x_2$, $x_2$
and skewness parameter
$\xi$.
Let us stress that our subsequent analysis is completely general: all invariant functions can be
treated at the same footing. For simplicity in what follows we employ the  same notation
for all the invariant functions
\be
H(x_1,\, x_2, \, x_3, \, \xi, \, t)\equiv \left\{V_i,\,A_i,\,T_i \right\}(x_1,\, x_2, \, x_3, \, \xi, \, u)\,.
\ee

A basic feature of model
building cleverness is to fulfill fundamental requirements of field theory, such as
general Lorentz covariance.
In particular this requirement leads to the so-called polynomiality property of the
Mellin moments in light-cone momentum fractions $x_1$, $x_2$, $x_3$ of $\pi N$ TDAs:
\be
&&
\int dx_1 dx_2 dx_3 \delta(2 \xi-x_1-x_2-x_3)  x_1^{n_1} x_2^{n_2} x_3^{n_3} H(x_1,x_2,x_3,\xi,u) \nonumber \\  &&
\sim \left. \left( i \frac{d}{dz_1^-} \right)^{n_1}
\left( i \frac{d}{dz_2^-} \right)^{n_2}
\left( i \frac{d}{dz_3^-} \right)^{n_3}
\left[ \langle \pi(P+\Delta/2) | \psi(z_1) \psi(z_2) \psi(z_3)|N(P-\Delta/2) \rangle \right] \right|_{z_{i}=0}\,.
\nonumber \\  &&
\label{polynomiality_ff}
\ee
Indeed the
$x_1$, $x_2$, $x_3$- Mellin moments of
$\pi N$ TDA are the form factors of the local twist-$3$ three quark operators
between nucleon and pion states.
This leads to the appearance of polynomials in
$\xi$
 at the right hand side of
(\ref{polynomiality_ff})%
\footnote{Naive counting gives $n_1+n_2+n_3$ for the order of this polynomial. However, the problem of determination of the highest possible power of $\xi$ in (\ref{polynomiality_ff}) still lacks some analysis.
This is a rather important question since it would allow to make the conclusion on the necessity of adding of $D$-term like
contributions \cite{Dterm} to
the spectral representation of $\pi N$ TDAs (see discussion in Sec.~\ref{Sec_Conclusions}). }.

%The first problem we are going to address is to specify the domain of definition of $\pi N$ TDAs.

% In our paper we mainly focus on the
%> $x_1$, $x_2$, $x_3$
%> and $\xi$
%> dependence of scalar form factors which arise at all possible
%> tensor structures occurring in the spin-flavor decomposition of
%> F.T. of matrix element of three-local light-cone quark operator
%> between meson and baryon states. Similarly as in the GPDs cases, each
%particuliar an axial or tensor
%> operators will certainly have a different set of independent
%> spin-flavor structures dotted by different scalar form factors.
%> However {\bf nearly} nothing would change in the dependence of these scalar
%> form factors on light cone momentum fractions
%> $x_i$ and $\xi$.
%> Indeed the spectral representation is a natural way to implement
%> the property of polynomiality of the Mellin moments which a consequence
%> of general Lorentz covariance. The only difference which may arise is
%> necessity to include certain $D$-term like contributions. The necessity
%> of including of such contributions for different $\pi N$ TDAs
%> requires however further investigation.

\section{Support properties of  $\pi N$ TDAs}
\label{Sec_Domains}

\subsection{ERBL-like and DGLAP-like domains for $\pi N$ TDAs}

\begin{figure}[h]
 \begin{center}
 \epsfig{figure=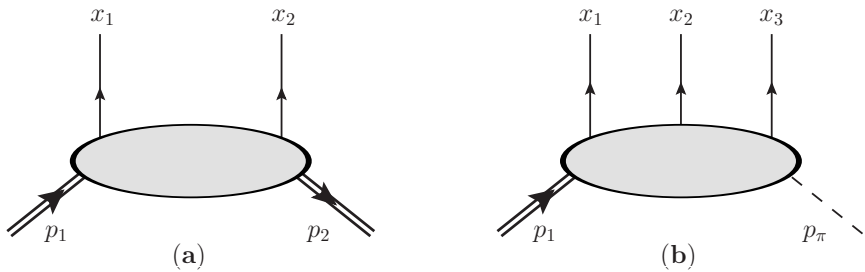 , height=3.5cm}
  \caption{Longitudinal momentum flow in the ERBL regime for GPDs~(a) and $\pi N$ TDAs~(b). }
\label{Blobs}
\end{center}
\end{figure}

In order to specify the support properties of
$\pi N$
TDAs let us first consider the case of the GPDs
(see Fig.~\ref{Blobs}.a).
Let
$x_1$
and
$x_2$
be the fractions
(defined with respect to average nucleon momentum
$P=\frac{p_1+p_2 }{2}$)
of the light-cone momentum carried by
quark and antiquark inside nucleon
($x_1+x_2=2 \xi$).
In the so-called ERBL region both
$x_1$
and
$x_2$
are positive. The variable
$x$
is usually defined as
\be
x=\frac{x_1-x_2}{2}
%{x_1+x_2}
\,.
\ee
In the ERBL region
$x_1, \,x_2 \in \,[0,\,2 \xi]$
and thus
$x \in \,[-\xi,\xi]$.
In the so-called DGLAP region either
$x_1$
is positive
$x_1 \in \,[2\xi,\,1+\xi]$
and $x_2$ is negative
$x_2 \in \,[-1+\xi,0]$ or
vice versa
($x_1$ is negative
$x_1 \in \,[-1+\xi,0]$ and
$x_2$
positive
$x_2 \in \,[2\xi,\,1+\xi]$).
These two  DGLAP domains result in
$x \in [\xi,\,1]$
and
$x \in [-1,\,-\xi]$
respectively.

Now let us turn to the case of
$\pi N$
TDAs.
Let $x_1$, $x_2$
and
$x_3$
satisfying the constraint
$x_1+x_2+x_3=2 \xi$,
with
$\xi \ge 0$
be the light-cone momentum fractions carried by
three quarks. As usual the light-cone momentum fractions are defined with respect to the average hadron momentum
$P=\frac{p_1+p_\pi }{2}$.
The convenient way to depict the support properties of
$\pi N$
TDAs is to employ barycentric coordinates (Mandelstam plane).

\begin{figure}[h]
 \begin{center}
 \epsfig{figure=  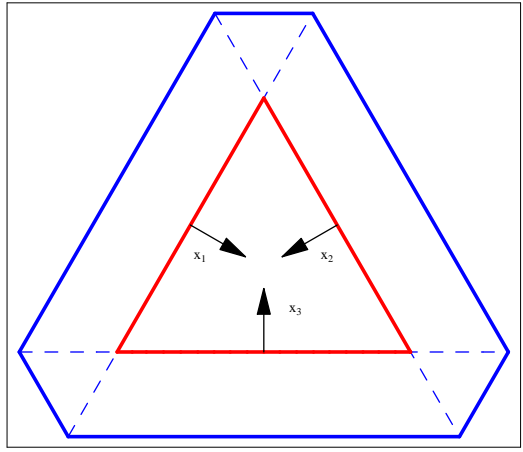 , height=6cm}
  \caption{Physical domains for $\pi N$ TDAs in the barycentric coordinates.}
\label{BaryCentric_Fig}
\end{center}
\end{figure}

First of all we identify the analogous of the ERBL domain,
in which three longitudinal momentum fraction carried by three quarks
are positive. In the barycentric coordinates the ERBL-like region corresponds
to the interior of the equilateral triangle with the height
$2 \xi$ (see Figure~\ref{BaryCentric_Fig}).
It is natural to assume that the DGLAP-like domains are bounded  by the lines
\be
&&
x_1=-1+ \xi\,; \ \  x_1=0 \,; \ \   x_1=1+\xi\,; \nonumber \\ &&
x_2=-1+ \xi\,; \ \  x_2=0 \,; \ \   x_2=1+\xi\,; \nonumber \\ &&
x_3=-1+ \xi\,; \ \  x_3=0 \,; \ \   x_3=1+\xi\,.
\label{Support_TDA_bary}
\ee
 We are guided by the following requirements.
\begin{itemize}
\item The complete domain of definition of
$\pi N$
TDA should be symmetric in
$x_1$, $x_2$, $x_3$.
\item In the limiting case
$\xi=1$
this domain should reduce to the ERBL-like domain on which
the nucleon DA is defined. In the barycentric coordinates
the domain of definition of the nucleon DA is equilateral triangle.
\item For any
$x_i$
set to zero we should recover the usual domain of definition of GPDs for the two remaining  variables.
\end{itemize}

Three small equilateral triangles correspond to DGLAP-like type I domains, where  only
one longitudinal momentum fractions is positive while two others are negative.
Three trapezoid domains correspond to DGLAP-like type II, where two longitudinal momentum fractions
are positive and one is negative.

\begin{figure}[h]
 \begin{center}
 \epsfig{figure=  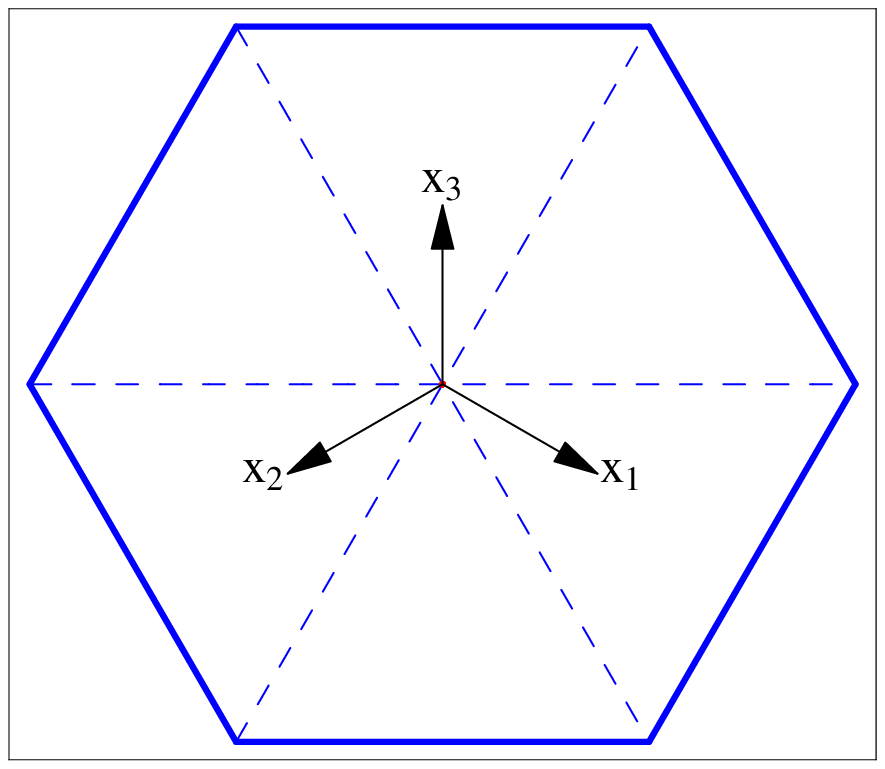 , height=4cm}
 \epsfig{figure=  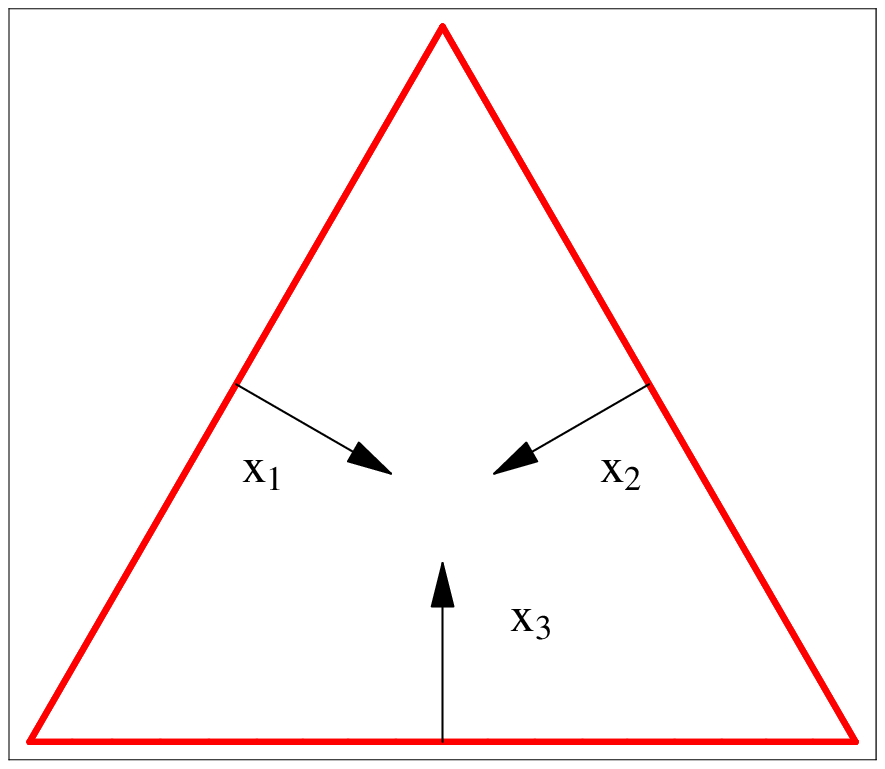 , height=4cm}
  \caption{Physical domains for $\pi N$ TDAs in the barycentric coordinates. Two limiting cases: $\xi=0$ (left) and $\xi=1$ (right). }
\label{Fig_domain_xi01}
\end{center}
\end{figure}

The support properties
(\ref{Support_TDA_bary})
are invariant under the permutation of the longitudinal momentum fractions $x_i$.
In the limit
$\xi \rightarrow 1$
the support of
$\pi N$ TDA is reduced to
the ERBL-like domain (the equilateral triangle)
(see Fig.~\ref{Fig_domain_xi01})
and coincide with that of the nucleon distribution amplitude (DA). In fact this is natural
since $\xi=1$ corresponds to the soft pion limit in which $\pi N$ TDA
reduces to the corresponding nucleon DA
\cite{Lansberg:2007ec}.

In the limiting case $\xi \rightarrow 0$
the support of
$\pi N$ TDA in the barycentric coordinates is given by the
regular hexagon.

\subsection{Quark-diquark coordinates}

In order to describe
$\pi N$ TDA instead of
$x_1$, $x_2$, $x_3$
which satisfy
\be
x_1+x_2+x_3=2 \xi\,,
\ee
it is convenient to introduce the so-called quark-diquark coordinates.
Let us stress that we do not imply any dynamical meaning to the notion of ``diquark''.
There are three different possible choices depending on which quarks
are supposed to form a ``diquark system'':
\be
&&
 v_{1}=\frac{x_2-x_3}{2}\,; \ \ \ w_{1}=\frac{x_1-x_2-x_3}{2}\,; \nonumber \\  &&
 v_{2}=\frac{x_3-x_1}{2}\,; \ \ \ w_{2}=\frac{x_2-x_3-x_1}{2}\,;  \nonumber \\  &&
 v_{3}=\frac{x_1-x_2}{2}\,; \ \ \ w_{3}=\frac{x_3-x_1-x_2}{2}\,.
\label{Quark-diquark_coordinates}
\ee
We  suggest  to introduce the notations
$\xi'_1$, $\xi'_2$ and $\xi'_3$
for the fraction of the longitudinal momentum
carried by the diquark:
\be
&&
\frac{x_2+x_3}{2}=  \frac{\xi-w_1}{2} \equiv  \xi'_1\,; \nonumber \\  &&
\frac{ x_1+x_3}{2} =\frac{\xi-w_2}{2} \equiv  \xi'_2\,; \nonumber \\  &&
\frac{x_1+x_2}{2}=  \frac{\xi-w_3}{2} \equiv   \xi'_3\,.
\label{def_xip}
\ee

The variables
$x_1$, $x_2$, $x_3$
are expressed through the new variables
(\ref{Quark-diquark_coordinates})
as follows:
\be
&&
x_1=\xi+w_1\,; \ \ \ x_2=v_1+\xi'_1\,; \ \ \ x_3=-v_1+\xi'_1\,; \nonumber \\ &&
%%%%%%%%%%%%%%%%
x_1=-v_2+\xi'_2\,; \ \ \ x_2=\xi+w_2\,; \ \ \ x_3= v_2+\xi'_2 \,; \nonumber \\ &&
x_1=v_3+ \xi'_3 %\frac{\xi-w}{2}
\, ; \ \ \ x_2=-v_3+ \xi'_3\,; \ \ \ x_3=\xi+w_3 \,.
\ee

\subsection{ ERBL-like and  DGLAP-like domains for $\pi N$ TDA in quark-diquark coordinates}

Let us consider how the ERBL-like and  DGLAP-like domains for $\pi N$ TDA look like in quark-diquark coordinates.
Throughout the rest of this section we employ the particular choice of
 quark-diquark coordinates
 (\ref{Quark-diquark_coordinates}):
\be
 v \equiv v_{3}=\frac{x_1-x_2}{2}\,; \ \ \ w\equiv w_{3}=\frac{x_3-x_1-x_2}{2}\,;
 \ \ \ \xi' \equiv \xi'_3 = \frac{\xi-w_3}{2}\,.
 \label{Quark-diquark_coordinates12}
\ee
The generalization for the alternative cases is straightforward.

\begin{figure}[h]
 \begin{center}
 \epsfig{figure=  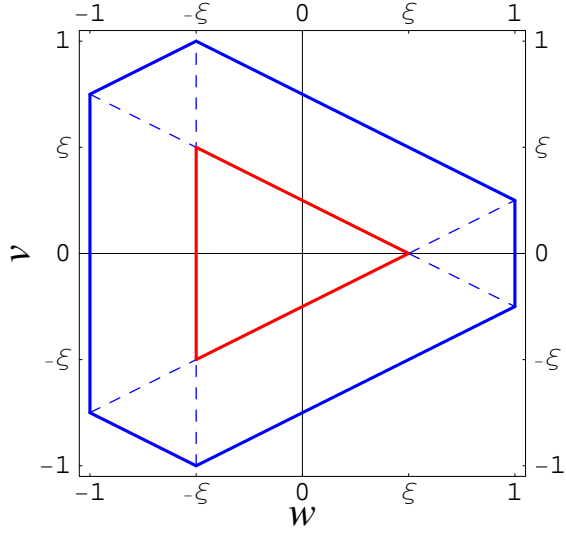 , height=7cm}
  \caption{ERBL-like and  DGLAP-like domains for $\pi N$ TDA in quark-diquark coordinates~(\ref{Quark-diquark_coordinates12}).
  Three lines: $w=-\xi$ and $v=\pm \xi'$
form the isosceles triangle which corresponds to ERBL-like region.  Three smaller isosceles triangles
correspond to DGLAP-like type I region.  Three trapezoid domains correspond to  DGLAP-like type II region.
   }
  \label{wv_Fig}
\end{center}
\end{figure}

The  ERBL-like and  DGLAP-like domains for
$\pi N$
TDA in quark-diquark coordinates
(\ref{Quark-diquark_coordinates12}).
are depicted on Figure~\ref{wv_Fig}.
In these coordinates the  ERBL-like region corresponds to the central isosceles triangular domain. It is
bounded by the lines
\be
v
%-\frac{\xi-w}{2}
=-\xi'\ \ (x_1=0)\,; \ \ \
v
%\frac{\xi-w}{2}
=\xi'\ \ (x_2=0)\,; \ \ \ w=-\xi \ \ (x_3=0)\,.
\ee
DGLAP-like  type I regions correspond to three smaller isosceles triangular domains.
Finally, three trapezoid domains correspond to  DGLAP-like type II region.

For $w \in
[-1,\;-\xi]$
DGLAP-like region is bounded by
\be
v=
%1+\frac{w+\xi}{2} \equiv
1+\xi-\xi' \ \ (x_1=1+\xi)  \ \  {\rm and} \ \
v=
%-1-\frac{w+\xi}{2} \equiv
-1-\xi+\xi' \ \ (x_2=1+\xi )\,.
\ee

For $w \in
[-\xi,\;\xi]$
DGLAP-like region is bounded by
\be
&&
v=
%-1+\frac{w+\xi}{2}=
-1+\xi-\xi' \ \ \ \   (x_1=-1+\xi) \,; \ \ \ v=-\xi'\,; \nonumber \\ &&
v= \xi'\,; \ \ \ \
v= %1-\frac{w+\xi}{2}=
1-\xi+\xi' \ \ (x_2=-1+\xi)\,.
\ee

For $w \in
[ \xi,\;1]$
DGLAP-like region is bounded by
\be
&&
v= %-1+\frac{w+\xi}{2}=
-1+\xi-\xi' \ \ \ \   (x_1=-1+\xi) \,; \ \ \ v=-\xi'\,; \nonumber \\ &&
v= \xi'\,; \ \ \ \
v=
%1-\frac{w+\xi}{2}=
1-\xi+\xi' \ \ (x_2=-1+\xi)\,.
\ee

One can easily check that for
$\xi \ge 0$
the following inequalities are valid:
\be
 \begin{cases} w<-\xi \\  \xi'> \xi \end{cases} ; \ \ \
{\rm and} \ \ \
\begin{cases} w>-\xi \\  \xi'<\xi \end{cases}.
\ee
$\xi'=\xi$
occurs on the line
$w=-\xi$.
Thus the whole domain of definition of
$\pi N$ TDA
in quark-diquark coordinates
depicted on Figure~\ref{wv_Fig}
can parameterized as follows:
\be
-1 \le w \le 1\,; \ \ \ \ -1+|\xi-\xi'| \le v \le 1- |\xi - \xi'|\,.
\label{The_whole_domain}
\ee
%where $\xi'$ is defined in (\ref{def_xip})\,.
% isosceles rather than equilateral triangle

Let us briefly summarize our result.
\begin{itemize}
\item $w \in \, [-1;\,-\xi]$   with
$v \in \, [ \xi';\, 1-\xi'+\xi]$
or
$v \in \, [ -1+\xi'-\xi;\, -\xi']$
correspond to DGLAP-like type I domains.

\item $w \in \, [-1;\,-\xi]$ and $v \in \, [-\xi';\, \xi']$ corresponds to  DGLAP-like type II domain.

\item  $w \in \, [-\xi;\, \xi]$ with $v \in \, [-\xi';\, \xi']$  corresponds to  ERBL-like domain.

\item  $w \in \, [-\xi;\, \xi]$ with $v \in \, [ \xi';\, 1-\xi+\xi']$
or
$v \in \, [ -1+\xi-\xi';\, -\xi']$
correspond to DGLAP-like type II domain.

\item  $w \in \, [\xi;\, 1]$ with $v \in \, [ -\xi';\, 1-\xi+\xi']$
or
$v \in \, [ -1+\xi-\xi';\, \xi']$
correspond to DGLAP-like type II domain.

\item  $w \in \, [\xi;\, 1]$ with $v \in \, [ \xi';\,  -\xi']$
correspond to DGLAP-like type I domain.

\end{itemize}

The Mellin moments of
$\pi N$
TDAs in
$x_1$, $x_2$, $x_3$
computed with the weight
\be
\int dx_1 dx_2 dx_3 \, \delta(x_1+x_2+x_3-2 \xi)
\ee
are the quantities of major theoretical importance.
In the quark-diquark coordinates
(\ref{Quark-diquark_coordinates12})
the corresponding integrals can be rewritten as
\be
&&
\int_{-1+\xi}^{1+\xi} dx_1 dx_2 dx_3 \,
%\int_{-1+\xi}^{1+\xi}
%\int_{-1+\xi}^{1+\xi}
\delta(x_1+x_2+x_3-2\xi)
x_1^{n_1}
x_2^{n_2}
x_3^{n_3}
H(x_1,\,x_2,\,x_3=2\xi-x_1-x_2)
\nonumber \\ && =
\int_{-1}^{1} dw \int_{-1+|\xi-\xi'|}^{1-|\xi-\xi'|} dv
(v+\xi')^{n_1}
(-v+\xi')^{n_2}
(w+\xi)^{n_3} H(w,\,v,\,\xi)\,.
\ee

\section{Spectral representation for $\pi N$ TDAs from the support properties and the polynomiality condition}
The double distribution representation
\cite{RDDA1,RDDA2,RDDA3,RDDA4}
was found to
be an elegant way to incorporate both the polynomiality property of the Mellin moments
and the support properties of GPDs.
In the framework of this representation the GPD
$H$
is given as a one dimensional section of the double distribution (DD)
$f(\alpha, \beta)$:
\be
H(x,\,\xi)= \int_{-1}^1 d \beta \int_{-1+|\beta|}^{1-|\beta|}
d \alpha \, \delta(x-\beta-\alpha \xi) f(\beta,\,\alpha)\,.
\label{DD_Rad}
\ee
The spectral representation
(\ref{DD_Rad})
was originally recovered in the diagrammatical
analysis employing the $\alpha$-representation techniques
\cite{Radyushkin:1983ea,Radyushkin:1983wh}.
The  spectral conditions
$|\beta| \le 1$
and
$|\alpha| \le 1-|\beta|$
ensure the support property of GPD
$|x| \le 1$ for any
$|\xi|\le 1$.

The polynomiality property of the Mellin moments in $x$
which resides on the fundamental field theoretic requirements
(Lorentz covariance) is ensured by the fact that
the $x$ dependence of GPD in
(\ref{DD_Rad})
is introduced solely through the integration path.
In
\cite{Teryaev:2001qm}
it was pointed out that the relation between GPDs
and DDs is the particular case of the Radon transform. It is worth
to mention that the polynomiality property is well known
in the framework of the Radon transform theory as the Cavalieri conditions
\cite{Gelfand_Graev}.

Now we propose to invert the logic.
From the pure mathematical point of view
representing GPD as the Radon transform
of a certain spectral density is the most natural way to ensure polynomiality property.
Postulating the polynomiality property of GPD and the support property
$|x| \le 1$
one can put down the spectral representation
(\ref{DD_Rad})
and unambiguously recover the spectral conditions
$|\beta| \le 1$
and
$|\alpha|\le 1-|\beta|$.
Let us stress that this does not provide the alternative derivation of
(\ref{DD_Rad})
since there is no way to show independently the support property
$|x| \le 1$
of GPD. However we think that this line of argumentation justifies
the use of the Radon transform
(\ref{DD_Rad})
which is a rather general representation for a function
satisfying the polynomiality condition with the restricted support in
$x$
as the building block for the spectral representation of multipartonic
generalizations of GPDs and in particular for
$\pi N$
TDAs.
In order to derive the form of the spectral representation for
$\pi N$
TDA let us first consider the simple example of ordinary GPDs.

\subsection{Test ground: spectral representation for GPDs}

We are going to treat the example of usual GPDs in a slightly unusual way which
we find more suitable for further generalization. Let us introduce
the light-cone momentum fractions
$x_1$
and
$x_2$
of the average hadron momentum carried by the quark and antiquark respectively.
The variables
$x_1$
and
$x_2$ satisfy the condition
$x_1+x_2=2\xi$.
The support property in
$x_1$,
$x_2$
is known to be given by
\be
-1+\xi \le x_1 \le 1+\xi\,; \ \ \ -1+\xi \le x_2 \le 1+\xi\,.
\label{support_prop_GPD}
\ee

In order to write down the spectral representation for GPD
we introduce two sets of spectral parameters
$\beta_{1,2}$, $\alpha_{1,2}$.
The momentum fractions
$x_{1,2}$
are supposed to have the following decomposition in terms of spectral parameters:
\be
x_1=\xi+\beta_1+\alpha_1 \xi\,; \ \ \ x_2=\xi+\beta_2+\alpha_2 \xi\,.
\ee
The condition
$x_1+x_2=2 \xi$
can be taken into account by introducing
two $\delta$-functions
$\delta(\beta_1+\beta_2) \delta(\alpha_1+\alpha_2)$.
This allows us  to write down the following spectral representation for GPD
$H(x_1,\,x_2=2 \xi-x_1,\,\xi)$:
\be
&&
H(x_1,\,x_2=2 \xi-x_1,\,\xi) \nonumber \\ && = \int_{\Omega_1} d \beta_1 d \alpha_1
\int_{\Omega_2} d \beta_2 d \alpha_2
\delta(x_1-\xi-\beta_1-\alpha_1 \xi)
\delta(\beta_1+\beta_2) \delta(\alpha_1+\alpha_2)
F(\beta_1, \beta_2, \alpha_1, \alpha_2)\,.
\nonumber \\ &&
\label{start_GPD0}
\ee
Here by
$\Omega_{1,2}$
we denote the usual domains in the parameter space:
\be
\Omega_{1,2}= \{ |\beta_{1,2}| \le 1\,; \ \ |\alpha_{1,2}| \le 1-|\beta_{1,2}|  \}\,;
\label{Rad_spectral}
\ee
and
$F(\beta_1, \beta_2, \alpha_1, \alpha_2)$
is a certain quadruple distribution.

The important advantage of the spectral representation
(\ref{start_GPD0})
is that it is symmetric under the interchange of the longitudinal
momentum fractions
$x_1$
and
$x_2$.
Note that the spectral conditions
(\ref{Rad_spectral})
ensure the support properties
(\ref{support_prop_GPD})
both in
$x_1$
and
$x_2$.
The
$(n_1,\,n_2)$-th
Mellin moments in
$x_1$, $x_2$
of
$H(x_1,\,x_2=2 \xi-x_1,\,\xi)$
are polynomials of order
$n_1+n_2$
of
$\xi$:
\be
&&
\nonumber
\int_{-1+\xi}^{1+\xi} dx_1 \int_{-1+\xi}^{1+\xi} dx_2  \,  \delta(2 \xi -x_1-x_2) x_1^{n_1}  x_2^{n_2}  H(x_1,\, x_2=2\xi-x_1,\,\xi) \\ &&
=\int_{\Omega_1} d \beta_1 d \alpha_1
\int_{\Omega_2} d \beta_2 d \alpha_2 \,
(\xi+\beta_1+\alpha_1 \xi)^{n_1}
(\xi+\beta_2+\alpha_2 \xi)^{n_2}   \nonumber \\ && \times
\delta(\beta_1+\beta_2) \delta(\alpha_1+\alpha_2)
F(\beta_1, \beta_2, \alpha_1, \alpha_2) =P_{n_1+n_2}(\xi)\,.
%\nonumber \\ &&
\ee

Now we are about to show that the spectral representation
(\ref{start_GPD0})
is equivalent to the usual Radyushkin's representation
(\ref{DD_Rad})
for GPDs in terms of double rather than quadruple
distributions. For this issue we can lift the two superfluous integrations
employing the two delta functions. In order to perform this in the
astute way let us introduce the natural  spectral variables
$\alpha_\pm$, $\beta_\pm$:
\be
\alpha_{\pm}= \frac{\alpha_1 \pm \alpha_2}{2}\,;  \ \ \  \beta_{\pm}= \frac{\beta_1 \pm \beta_2}{2}\,.
\label{natural_variables_ab}
\ee
It is also useful to
to perform the related change of the variables in the
$(x_1, \, x_2)$
space in the initial spectral representation
(\ref{start_GPD0}).
The corresponding natural variables are
\be
x_-= \frac{x_1-x_2}{2}=\alpha_-+ \beta_-\xi\,; \ \ \ \text{and} \ \ \ x_+= \frac{x_1+x_2}{2}= \xi+ \alpha_++ \beta_+\xi\,.
\ee
Thus instead of using
(\ref{start_GPD0})
we switch to the natural variables and consider:
\be
&&
H(x_1,\,x_2= 2\xi-x_1,\,\xi) \nonumber \\ && = \frac{1}{2} \int_{\Omega_1} d \beta_1 d \alpha_1
\int_{\Omega_2} d \beta_2 d \alpha_2
\delta(x_--\beta_--\alpha_- \xi)
\delta(\beta_+) \delta(\alpha_+)
F(\beta_1, \beta_2, \alpha_1, \alpha_2)\,.
\nonumber \\ &&
\label{Spectral_for_GPDs_x12_refined}
\ee

The appropriate definition of the integration domain in
(\ref{Spectral_for_GPDs_x12_refined})
after the change of the variables
(\ref{natural_variables_ab})
require special attention.
In particular,
\be
\int_{-1}^1 d \beta_1 \int_{-1}^1 d \beta_2 \, ... = 2 \int_{-1}^1 d \beta_- \int_{-1+|\beta_-|}^ {1-|\beta_-|} d \beta_+ \,...\;.
\ee
Now since
$1-|\beta_-| \ge 0$
and hence
$-1+|\beta_-| \le 0$
the integral over
$\beta_+$
can be easily lifted with no influence on the integration domain in
$\alpha_+$,
$\alpha_-$.
The  problem of definition of the integration domain in $\alpha_+$,
$\alpha_-$  in principle is reduced to change of the variables in the
integral
\be
\int_{-a}^{a} d \alpha_1 \int_{-b}^{b} d \alpha_2 \delta(\alpha_1+\alpha_2)\, ... \,,
\ee
where
$a=1-|\beta_++\beta_-|$,
$b=1-|\beta_+-\beta_-|$.
It is much simplified due to the fact that $\beta_+=0$ and thus $a=b \equiv 1-|\beta_-|$.
This gives
\be
\int_{- a}^{ a} d \alpha_1 \int_{-a}^{a} d \alpha_2 \delta(\alpha_1+\alpha_2)\, ... =
2 \int_{- a}^{ a} d \alpha_- \int_{-a+|\alpha_-|}^{a-|\alpha_-|} d \alpha_+ \,  \delta(\alpha_+)...\;.
\ee
Now the integral over $\alpha_+$ can be trivially performed with the help of $\delta$-function
again producing no additional restrictions for the integration domain in
$\alpha_-$
and
$\beta_-$.
The final result reads
\be
&&
H(x_1,\,x_2=2\xi-x_1,\,\xi)\nonumber \\ && =
\int_{-1}^1 d \beta_- \int_{-1+|\beta_-|}^{1-|\beta_-|} d \alpha_-
\delta(x_--\beta_--\alpha_-\xi)
\underbrace{2 F \left( \beta_-, -\beta_-, \alpha_-,-\alpha_- \right)}_{f(\beta_-, \, \alpha_-)}\,.
\ee
Certainly we just recovered the known Radyushkin's result for the double distribution representation
of GPDs.

Let us just make a short summary of the crucial points.
\begin{itemize}
\item We started from the spectral representation
for
$H(x_1,\,x_2=2\xi-x_1, \xi)$
as the function of the skewness parameter
$\xi$
and of two longitudinal momentum fractions
$x_1$, $x_2$
satisfying the condition
$x_1+x_2=2\xi$.
The form of this spectral representation ensured the proper support properties in
$x_1$, $x_2$
as well as the polynomiality property of the corresponding Mellin moments in
$x_1$ and $x_2$.
The spectral density was a certain quadruple rather than double distribution.

\item The constraint
$x_1 + x_2= 2 \xi$
was taken into account by the introduction of
two $\delta$-functions
restricting the integration domain in the space of spectral variables.

\item The two superfluous integrations can be lifted with the help of two $\delta$-functions. This
requires the special attention to the integration domain in the space of spectral parameters.
This problem can be most easily solved by switching to the set of natural variables {\emph both} in the
space of spectral parameters and
$x_1$, $x_2$
space.

\item In our toy exercise lifting the two integrations does not
lead to any special restrictions on the remaining spectral parameters
$\alpha_-$, $\beta_-$ and we just recover the usual Radyushkin's result for the double distribution representation
of GPDs.

\item  We find the spectral representation
(\ref{start_GPD0})
which is symmetric under the exchange of $x_1$ and $x_2$
suitable for the generalization to the multiparton case.
 The analysis of $\pi N$ TDAs with the help of the approach discussed above is presented in the
next subsection.
\end{itemize}

\subsection{Spectral representation for $\pi N$ TDAs}

We are now about to apply the ideas described in the previous section to the case of
$\pi N$
TDAs.
Let us consider
$\pi N$
TDA
$H(x_1,\,x_2, \, x_3=2 \xi-x_1-x_2, \,\xi)$
as a function of light-cone momentum fractions
$x_1$, $x_2$
and
$x_3$
carried by three quarks. The three light-cone momentum fractions
satisfy the condition
$x_1+x_2+x_3=2 \xi$.
The support property in
$x_1$,
$x_2$,
$x_3$
is given by
\be
-1+\xi \le x_1 \le 1+\xi\,; \ \ \ -1+\xi \le x_2 \le 1+\xi\,; \ \ \ -1+\xi \le x_3 \le 1+\xi\,.
\label{support_prop_PiNTDA_II}
\ee

In order to write down the spectral representation for
$H(x_1,\,x_2, \, x_3=2 \xi-x_1-x_2, \,\xi)$
we introduce three sets of spectral parameters
$\beta_{1,2,3}$, $\alpha_{1,2,3}$.
The momentum fractions
$x_{1,2,3}$
are supposed to have the following decomposition in terms of spectral parameters:
\be
x_1=\xi+\beta_1+\alpha_1 \xi\,; \ \ \ x_2=\xi+\beta_2+\alpha_2 \xi\,; \ \ \ x_3=\xi+\beta_3+\alpha_3 \xi\,.
\ee
In order to satisfy this constrain
we  require that
\be
\beta_1+ \beta_2+ \beta_3=0\, ; \ \ \ \alpha_1+\alpha_2+\alpha_3=-1\,.
\ee

This allows to write down
the following spectral representation for $\pi N$ TDAs:
%$H(x_1,\,x_2,\,x_3=2 \xi -x_1-x_2,\,\xi)$:
\be
&&
H(x_1,\,x_2,\,x_3=2 \xi -x_1-x_2,\,\xi) \nonumber \\ && =
\left[
\prod_{i=1}^3
\int_{\Omega_i} d \beta_i d \alpha_i
\right]
%\int_{\Omega_2} d \beta_2 d \alpha_2
%\int_{\Omega_3} d \beta_3 d \alpha_3
\delta(x_1-\xi-\beta_1-\alpha_1 \xi) \,
\delta(x_2-\xi-\beta_2-\alpha_2 \xi) \,
%\delta(x_3-\xi-\beta_3-\alpha_3 \xi) \,
\nonumber \\ &&
\times
\delta(\beta_1+ \beta_2+ \beta_3)
\delta(\alpha_1+\alpha_2+\alpha_3+1)
 F(\beta_1, \, \beta_2, \, \beta_3, \, \alpha_1, \, \alpha_2, \alpha_3)\,.
%\nonumber \\ &&
\label{Spectral_for_GPDs_x123}
\ee
By
$\Omega_{i},\,i= \left\{ 1, \, 2, \,3 \right\}$
we denote the usual domains in the parameter space:
\be
\Omega_{i}= \{ |\beta_{i}| \le 1\,; \ \ |\alpha_{i}| \le 1-|\beta_{i}|  \}\,;
\label{Rad_spectral_II}
\ee
and $F(\beta_1, \, \beta_2, \, \beta_3, \, \alpha_1, \, \alpha_2, \alpha_3)$
is now a sextuple  distribution.
The spectral conditions
(\ref{Rad_spectral_II})
ensure the support properties
(\ref{support_prop_PiNTDA_II}).
Obviously, the
$(n_1,\, n_2, \, n_3)$-th Mellin moment in
$(x_1,\, x_2, \, x_3)$
of
$\pi N$
TDA is a polynomial of order
$n_1+n_2+n_3$
of
$\xi$:
\be
&&
\left[
\prod_{i=1}^3
\int_{-1+\xi}^{1+\xi} d x_i
\right] \, \delta(x_1+x_2+x_3-2 \xi)
x_1^{n_1} x_2^{n_2} x_3^{n_3}
H(x_1,\,x_2,\,x_3=2 \xi -x_1-x_2,\,\xi)
\nonumber \\ && = P_{n_1+n_2+n_3}(\xi)\,.
\ee

In complete analogy with the previously considered case of usual GPDs
in order to properly reduce the spectral representation in terms of sextuple
distribution for $\pi N$ TDA to that in terms of quadruple  distribution
we need to perform two integrations in
\be
\left[
\prod_{i=1}^3
\int_{\Omega_i} d \beta_i d \alpha_i
\right] \delta(\beta_1+ \beta_2+ \beta_3)  \delta(\alpha_1+\alpha_2+\alpha_3+1) \,...
\ee
employing $\delta$-functions and specify the integration limits in the remaining four integrals.
This problem can be solved by introducing the appropriate natural variables.

Let us start with the integral
\be
\int_{-1}^1 d \beta_1 \int_{-1}^1 d \beta_2 \int_{-1}^1 d \beta_3  \,
\delta(\beta_1+ \beta_2+ \beta_3)\,.
\label{Constrained_Int_beta_initial}
\ee
In order to visualize the integration domain
(\ref{Constrained_Int_beta_initial})
it is natural to employ the barycentric coordinates.
In these coordinates the domain selected by the conditions
$|\beta_i| \le 1$ ($i \in \{1,\,2,\,3 \}$)
and
$\beta_1+\beta_2+\beta_3=0$
is represented by a regular hexagon (confer Fig~\ref{Fig_domain_xi01}).
It is convenient to single out three domains inside this hexagon:
\be
&&
D_1: \ \ \ \{ \beta_1 \ge 0, \,  \beta_2 \le 0, \, \beta_3 \le 0 \}   \cup  \{ \beta_1 \le 0, \,  \beta_2 \ge 0, \, \beta_3 \ge 0 \}\,;  \nonumber \\ &&
D_2:  \ \ \ \{ \beta_2 \ge 0, \,  \beta_1 \le 0, \, \beta_3 \le 0 \}   \cup  \{ \beta_2 \le 0, \,  \beta_1 \ge 0, \, \beta_3 \ge 0 \}\,;  \nonumber \\ &&
D_3: \ \ \  \{ \beta_3 \ge 0, \,  \beta_1 \le 0, \, \beta_2 \le 0 \}   \cup  \{ \beta_3 \le 0, \,  \beta_1 \ge 0, \, \beta_2 \ge 0 \}\,.
\label{Def_3_Domains}
\ee
Obviously
\be
\int_{-1}^1 d \beta_1 \int_{-1}^1 d \beta_2 \int_{-1}^1 d \beta_3  \,
\delta(\beta_1+ \beta_2+ \beta_3)=
\sum_{i=1}^3 \int_{D_i} d \beta_1 d \beta_2  d \beta_3 \, \delta(\beta_1+ \beta_2+ \beta_3)\,.
 \label{Constrained_Int_beta}
\ee

Now in order to get rid of one of three integrations in
(\ref{Constrained_Int_beta})
we should switch to the natural coordinates.
There are three possible choices of  the natural coordinates in
(\ref{Constrained_Int_beta}).
For the moment we are going to adopt the coordinates
\be
\rho_3= \frac{\beta_1-\beta_2}{2}\,; \ \ \ \sigma_3= \frac{\beta_3-\beta_1-\beta_2}{2}\,.
\label{Nat_beta_3}
\ee
The constrained triple integral
(\ref{Constrained_Int_beta_initial})
can be then rewritten as
\be
\int_{-1}^1 d \sigma_3 \int_{-1+\frac{|\sigma_3|}{2}}^{1-\frac{|\sigma_3|}{2}} d \rho_3\,...\,.
\ee
In principle in a completely analogous way one may also employ the coordinates
\be
&&
\rho_1= \frac{\beta_2-\beta_3}{2}\,; \ \ \ \sigma_1= \frac{\beta_1-\beta_2-\beta_3}{2}\,; \nonumber \\ &&
\rho_2= \frac{\beta_3-\beta_1}{2}\,; \ \ \ \sigma_2= \frac{\beta_2-\beta_3-\beta_1}{2}
\ee
yielding the result
\be
\int_{-1}^1 d \sigma_i \int_{-1+\frac{|\sigma_i|}{2}}^{1-\frac{|\sigma_i|}{2}} d \rho_i\,...\,.
\ee

%\subsection*{A technical problem we need to solve}
Now let us address the problem of computation
of the constrained triple integral over $\alpha_i$ in
(\ref{Spectral_for_GPDs_x123}):
\be
\int_{-a}^a d \alpha_1
\int_{-b}^b d \alpha_2
\int_{-c}^c d \alpha_3
\, \delta(\alpha_1+\alpha_2+\alpha_3+1)\,...\,,
\label{Constrained_int_alpha}
\ee
where we introduced the notations
\be
a \equiv 1- |\beta_1|\,; \ \ \ b \equiv 1- |\beta_2|\,; \ \ \ c \equiv 1- |\beta_3|\,;
\ee
( $a \ge 0\,; \, a \le 1$,  $b \ge 0\,; \, b \le 1$,  $c \ge 0\,; \, c \le 1$).

Introducing the natural coordinates%
\footnote{There are two additional possible choices: $\omega_1=\alpha_1\,;\; \nu_1=\frac{\alpha_2-\alpha_3}{2}$
and $\omega_2=\alpha_2\,; \; \nu_2=\frac{\alpha_3-\alpha_1}{2}$. }
\be
\omega_3=\alpha_3\,; \ \ \  \nu_3=\frac{\alpha_1-\alpha_2}{2}
\label{Nat_alpha_3}
\ee
and employing the results of the
Appendix~\ref{App_Useful_int}
we  conclude that for
$\beta_i \in D_1 \cup D_2 \cup D_3$
the constrained integral
(\ref{Constrained_int_alpha})
can be rewritten as
\be
\int_{-1+|\beta_3|}^{1-|\beta_1|-|\beta_2|} d \omega_3
\int_{-1+|\beta_1|+\frac{1+\omega_3}{2}}^{ 1-|\beta_2|-\frac{1+\omega_3}{2}} d \nu_3   \, ... \,.
\ee

Now let us put all together and write down the spectral representation for
$\pi N$
TDAs in terms of quadruple distributions.
The important observation is that once
we have chosen the variables
$\sigma_3$,
$\rho_3$
and
$\omega_3$,
$\nu_3$
to perform the constrained integration in
$\beta_1$, $\beta_2$, $\beta_3$
and
$\alpha_1$, $\alpha_2$, $\alpha_3$
respectively
the natural variables on which
$\pi N$ TDAs depends are
\be
w_3= \frac{x_3-x_1-x_2}{2}, \ \ \ v_3=\frac{x_1-x_2}{2}\,.
\label{QdQ_3}
\ee
Expressing the
$\beta_i$ and $\alpha_i$
through
 $\sigma_3$, $\rho_3$, $\omega_3$, $\nu_3$
the two delta functions in the definition
(\ref{Spectral_for_GPDs_x123})
can be traded for
\be
&&
\nonumber
\left. \delta(x_1-\xi-\beta_1-\alpha_1 \xi) \, \delta(x_2-\xi-\beta_2-\alpha_2 \xi)
\right|_{x_1+x_2+x_3=2 \xi} \\ && =
\delta(w_3-\sigma_3- \omega_3 \xi) \, \delta(v_3-\rho_3-\nu_3 \xi)\,.
\ee
Note that at the level of
delta functions
we achieved the ``factorization'' of
$w_3$ and $v_3$ dependencies on the spectral parameters.

Thus in the natural spectral parameters
(\ref{Nat_beta_3}), (\ref{Nat_alpha_3})
and quark-diquark coordinates
(\ref{QdQ_3})
we recovered the form of the spectral representation of
$\pi N$
TDAs in terms of quadruple distributions:
\be
&&
H
%_3
(w_3,\,v_3,\,\xi) \nonumber \\ && = \int_{-1}^1 d \beta_1 d \beta_2 d \beta_3 \,
\delta(\beta_1+\beta_2+\beta_3)
\int_{-1+|\beta_1|}^{1-|\beta_1|} d \alpha_1
\int_{-1+|\beta_2|}^{1-|\beta_2|} d \alpha_2
\int_{-1+|\beta_3|}^{1-|\beta_3|} d \alpha_3
\delta(\alpha_1+\alpha_2+\alpha_3+1)
\nonumber \\ &&
\delta(x_1-\xi-\beta_1-\alpha_1 \xi) \,
\delta(x_2-\xi-\beta_2-\alpha_2 \xi) \,
F(\beta_1,\, \beta_2,\, \beta_3,\, \alpha_1,\, \alpha_2, \alpha_3)
\nonumber \\ &&
= \int_{-1}^1 d \sigma_3
\int_{-1+\frac{|\sigma_3|}{2}}^{1-\frac{|\sigma_3|}{2}} d \rho_3
\int_{-1+|\sigma_3|}^{1-|\rho_3- \frac{\sigma_3}{2} |-|\rho_3+ \frac{\sigma_3}{2}|} d \omega_3
\int_{-\frac{1}{2}+|\rho_3- \frac{\sigma_3}{2}|+\frac{ \omega_3}{2}}^{\frac{1}{2}-|\rho_3+ \frac{\sigma_3}{2}|-\frac{ \omega_3}{2}} d \nu_3
\delta(w_3-\sigma_3-\omega_3 \xi)
\nonumber \\ &&
\times  \delta(v_3-\rho_3-\nu_3 \xi) \,
F_3(\sigma_3,\, \rho_3,\, \omega_3,\, \nu_3)\,,
\label{spectral_representation_D1}
\ee
where
\be
F_3(\sigma_3,\, \rho_3,\, \omega_3,\, \nu_3) \equiv
F(\rho_3-\frac{\sigma_3}{2},\, -\rho_3-\frac{\sigma_3}{2},\, \sigma_3,\, \nu_3- \frac{1+\omega_3}{2},\, -\nu_3- \frac{1+\omega_3}{2}, \omega_3)\,.
\label{Def_F3}
\ee

Employing three possible sets of natural spectral parameters
one can write down three equivalent spectral representations
in terms of three sets of quark-diquark coordinates $w_i$,
$v_i$
with
$i= 1,\,2,\,3 $
defined in
(\ref{Quark-diquark_coordinates}):
\be
&&
%\left. H_i(x_1,\,x_2,\,x_3,\,\xi) \right|_{x_1+x_2+x_3=2 \xi}
H%_i
(w_i,\,v_i,\,\xi)
\nonumber \\ &&
=
\int_{-1}^1 d \sigma_i
\int_{-1+\frac{|\sigma_i|}{2}}^{1-\frac{|\sigma_i|}{2}} d \rho_i
\int_{-1+|\sigma_i|}^{1-| \rho_i- \frac{\sigma_i}{2}|-|\rho_i+ \frac{\sigma_i}{2}|} d \omega_i
\int_{-\frac{1}{2}+|\rho_i- \frac{\sigma_i}{2}|+\frac{\omega_i}{2}}^{\frac{1}{2}-|\rho_i+ \frac{\sigma_i}{2}|-\frac{ \omega_i}{2}} d \nu_i
\delta(w_i-\sigma_i-\omega_i \xi)
\nonumber \\ &&
\times  \delta(v_i-\rho_i-\nu_i \xi) \,
 F_i(\sigma_i,\, \rho_i,\, \omega_i,\, \nu_i)\,,
\label{Spectral_represent_Hi}
\ee
where
$F_3(\sigma_3,\, \rho_3,\, \omega_3,\, \nu_3)$
is defined in
(\ref{Def_F3}) and
\be
&&
F_1(\sigma_1,\, \rho_1,\, \omega_1,\, \nu_1) \equiv F(\sigma_1, \,\rho_1-\frac{\sigma_1}{2}\,,-\rho_1-\frac{\sigma_1}{2},\,
\omega_1,\, \nu_1-\frac{1+\omega_1}{2}, \, -\nu_1-\frac{1+\omega_1}{2})\,;
\nonumber \\ &&
F_2(\sigma_2,\, \rho_2,\, \omega_2,\, \nu_2)
\equiv F(-\rho_2-\frac{\sigma_2}{2},\, \sigma_2,\,\rho_2-\frac{\sigma_2}{2},\,
 -\nu_2-\frac{1+\omega_2}{2},\,\omega_2,\,  \nu_2-\frac{1+\omega_2}{2})\,.
 \nonumber \\ &&
\ee
%In principle there is no reason to prefer any particular form
%(\ref{Spectral_represent_Hi}).
%Instead it may turn to be useful to present $\pi N$ TDA in
%a symmetric form as a sum of three items:
%\be
%\left. H_i(x_1,\,x_2,\,x_3,\,\xi) \right|_{x_1+x_2+x_3=2 \xi} %\nonumber \\ &&
%=  \sum_{i=1}^3 H_i(w_i,\,v_i,\,\xi)\,.
%\label{Spectral_represent_SUM_OF_THREE}
%\ee

The spectral representation
(\ref{Spectral_represent_Hi})
for $\pi N$ TDA in terms of quadruple distribution
is the main result of our paper.
However this form of the result is still not very convenient for practical applications.
In the next section we demonstrate that the spectral representation
(\ref{Spectral_represent_Hi})
satisfies the support properties of $\pi N$ TDAs
established in Sec.~\ref{Sec_Domains}. We also derive the explicit expressions for
 $\pi N$ TDAs in the ERBL-like and DGLAP-like type I and II domains.

\section{Support properties of $\pi N$ TDAs and the spectral representation}

In order to make our formulas more compact in what follows we omit the indice $i$ for
the quark-diquark coordinates $w_i$ and $v_i$, spectral parameters
$\sigma_i$, $\rho_i$, $\omega_i$, $\nu_i$ and the spectral densities $F_i$.
Our subsequent analysis equally applies for all
$i=1,\,2,\,3$.

It is extremely instructive to check that each contribution
into
$\pi N$ TDA
in
(\ref{Spectral_represent_Hi})
satisfies the support properties
which were established in Sec.~\ref{Sec_Domains}:
\be
-1 \le w \le 1\,; \ \ \ \ -1+|\xi- \xi'| \le v \le 1- |\xi - \xi'|\,
\label{The_whole_domain_i}
\ee
with
$\xi'$
defined in
(\ref{def_xip}).
In particular this allows to check that
$(N-n,n)$-th
($N \ge n \ge 0$)
Mellin moments of
$\pi N$
TDA in
$(w,v)$
indeed satisfy the polynomiality property:
\be
\int_{-1}^1 d w \int_{-1+|\xi- \xi'|}^{1-|\xi- \xi'|} d v \,
w^{N-n} v^n \, H(w,\,v,\,\xi)= P_N(\xi)\,,
\ee
where
$P_N(\xi)$
is a polynomial of order
$N$
in
$\xi$.

\subsection*{Case $\xi=0$}
Let us first consider the case
$\xi=0$.
Employing the first delta function
we get
$\sigma=w$
for
$-1\le w \le 1$
and
$0$ otherwise.
This obviously ensures the first condition
(\ref{The_whole_domain_i}) for
$\xi=0$.
Once the integral over $\sigma$ is performed the dependence on
$v$
is  introduced through
\be
\int_{-1+\frac{|w |}{2}}^{1-\frac{|w |}{2}} d \rho  \, \delta(v -\rho )\,...\,.
\ee
The result of this integral is non-zero only for
\be
-1+\frac{|w |}{2} \le v  \le 1- \frac{|w |}{2}\,,
\ee
that is precisely the second condition
(\ref{The_whole_domain_i})
for
$\xi=0$.

\subsection*{Case $0< \xi \le 1$}
Let us now show that the spectral representation
(\ref{spectral_representation_D1})
possesses the desired support properties for arbitrary value of $\xi \in (0;\,1]$%
\footnote{The final result for $\xi \in [-1;\,0)$ is presented in the Appendix~\ref{App_xi_le_0}.}.

First of all it is easy to see that the first one of the two conditions
(\ref{The_whole_domain_i})
is respected. Indeed the
$w$
dependence in
(\ref{Spectral_represent_Hi})
is introduced through the  expression
\be
 \int_{-1}^1 d \sigma  \int_{-1+|\sigma |}^{1-| \rho - \frac{\sigma }{2}| -| \rho + \frac{\sigma }{2}|} d \omega \, \delta(w-\sigma-\omega \xi)\,...\,.
\label{w_dependece_introduced}
\ee
From the inequalities
(\ref{Ineq_1}),
(\ref{Ineq_2})
and
(\ref{Ineq_3})
it follows that
\be
-1+|\sigma| \le 1-| \rho- \frac{\sigma}{2}| -| \rho+ \frac{\sigma}{2}| \le 1-|\sigma|\,.
\ee
Thus in
(\ref{w_dependece_introduced})
we are integrating only over some part of the familiar ``GPD square''
$|\rho| \le 1-|\sigma|$. This guarantees the vanishing of $\pi N$ TDA for
$|w|>1$.
One can in the usual way perform the integration
over
$\omega$
introducing the additional
$\theta$-function to take into the account the unusual
upper limit in the
integral over
$\omega$:
\be
\theta( 1-|\rho - \frac{\sigma }{2}|-|\rho + \frac{\sigma }{2}|-\frac{w-\sigma}{\xi}) \equiv \theta(...)\,.
\ee

For
$\xi > 0$
we get
\be
&& H(w,\,v,\,\xi)=   \nonumber \\ &&
\left[
\begin{split} &
\text{For}  \ \  w  \in ( -\infty; \,  -1): \ \ 0\,;  \\   &
\text{For} \ \   w  \in [ -1; \,  -\xi]: \\   &
 \frac{1}{\xi }   \int_{\frac{w -\xi}{1+\xi}}^{\frac{w +\xi}{1-\xi}} d \sigma
\int_{-1+\frac{|\sigma |}{2}}^{1-\frac{|\sigma |}{2}} d \rho
\int_{-\frac{1}{2}+|\rho- \frac{\sigma }{2}|+\frac{w -\sigma }{2 \xi}}^{\frac{1}{2}-|\rho + \frac{\sigma }{2}|-\frac{w -\sigma }{2\xi}} d \nu \,
  \delta(v -\rho -\nu  \xi) \, \theta(...) \, F(\sigma , \rho , \frac{w -\sigma }{\xi}, \nu  )\,;
 \\   &
\text{For} \ \  w  \in [ -\xi; \,  \xi]: \\   &
 \frac{1}{\xi }   \int_{\frac{w -\xi}{1+\xi}}^{\frac{w +\xi}{1+\xi}} d \sigma
\int_{-1+\frac{|\sigma |}{2}}^{1-\frac{|\sigma |}{2}} d \rho
\int_{-\frac{1}{2}+|\rho - \frac{\sigma }{2}|+\frac{w -\sigma }{2 \xi}}^{\frac{1}{2}-|\rho + \frac{\sigma }{2}|-\frac{w -\sigma }{2\xi}} d \nu \,
  \delta(v -\rho -\nu  \xi) \, \theta(...)\, F(\sigma , \rho , \frac{w -\sigma }{\xi}, \nu  )\,;
  \\   &
\text{For}  \ \  w  \in [  \xi; \,  1]:  \\   &
\frac{1}{\xi }   \int_{\frac{w -\xi}{1-\xi}}^{\frac{w +\xi}{1+\xi}} d \sigma
\int_{-1+\frac{|\sigma |}{2}}^{1-\frac{|\sigma |}{2}} d \rho
\int_{-\frac{1}{2}+|\rho - \frac{\sigma }{2}|+\frac{w -\sigma }{2 \xi}}^{\frac{1}{2}-|\rho + \frac{\sigma }{2}|-\frac{w -\sigma }{2\xi}} d \nu \,
  \delta(v -\rho -\nu  \xi)  \, \theta(...) \, F(\sigma , \rho , \frac{w -\sigma }{\xi}, \nu  )\,;
  \\   &
\text{For}  \ \  w  \in ( 1; \,  \infty ): \ \ 0\,.
 \end{split}
 \right.
 \nonumber \\ &&
 \label{Omega_int_performed}
\ee

Now we are about to perform the integration over
$\nu $
with the help of the last remaining
$\delta$-function.
The resulting domain of integration in
$\sigma $
and
$\rho $
is defined by the inequalities
\be
&&
-1 +\frac{|\sigma| }{2} \le \rho  \le  1 -\frac{|\sigma| }{2}\,;
\label{Ineq_master1}
\ee
\be
-\frac{1}{2} + |\rho - \frac{\sigma }{2}| + \frac{w -\sigma }{2 \xi}\le \frac{v -\rho }{\xi}
\le  \frac{1}{2} - |\rho + \frac{\sigma }{2}| - \frac{w -\sigma }{2 \xi}\,;
\label{Ineq_master2}
\ee
\be
1-|\rho - \frac{\sigma }{2}|-|\rho + \frac{\sigma }{2}| \ge \frac{w-\sigma}{\xi}\,;
\label{Ineq_master3}
\ee
as well as the integration limits in
$\sigma $
depending on the value of
$w$ (see (\ref{Omega_int_performed})).

It can be shown that for
$\xi \ge 0$
the  two inequalities
(\ref{Ineq_master2})
are equivalent to
\be
&&
\rho  \le \frac{\sigma }{2}+ \frac{v + \xi' }{1+\xi} \ \ \text{for} \ \ v  \ge -\xi'\,; \ \ \ \  %\nonumber \\ &&
\rho  \le \frac{\sigma }{2}+ \frac{v + \xi' }{1-\xi} \ \ \text{for} \ \ v  \le -\xi'
\label{Ineq_master2_refined_1}
\ee
together with
\be
\rho  \ge -\frac{\sigma }{2}+ \frac{v -\xi' }{1-\xi} \ \ \text{for} \ \ v  \ge  \xi'  \,; \ \ \ \  %\nonumber \\ &&
\rho  \ge -\frac{\sigma }{2}+ \frac{v -\xi' }{1+\xi} \ \ \text{for} \ \ v  \le  \xi'\,.
\label{Ineq_master2_refined_2}
\ee

Analogously
the  inequality
(\ref{Ineq_master3})
for
$\xi \ge 0$
is equivalent to
\be
&& \rho \le \frac{\sigma}{2 \xi}+ \frac{\xi'}{\xi} \ \ \ \text{for} \ \ \ v \ge   |\xi'| \,; \ \ \ \  %\nonumber \\ &&
\rho \ge -\frac{\sigma}{2 \xi}- \frac{\xi'}{\xi} \ \ \ \text{for} \ \ \ v  \le  -|\xi'|  \,; \nonumber \\ &&
 \sigma \ge \frac{w-\xi}{1+\xi} \ \ \   \text{for} \ \ \  \begin{cases} v \ge -\xi' \\ v \le \xi' \end{cases} \,;   \ \ \ \  %\nonumber \\ &&
 \sigma \ge \frac{w-\xi}{1-\xi}  \ \ \  \text{for} \ \ \  \begin{cases} v \le -\xi' \\ v \ge \xi' \end{cases} \,.
\label{Ineq_master3_refined}
\ee

The last step is to match the integration domain defined by the inequalities
(\ref{Ineq_master1}),
(\ref{Ineq_master2_refined_1}),
(\ref{Ineq_master2_refined_2})
and
(\ref{Ineq_master3_refined})
with the explicit
$w $-dependent limits of integration in
$\sigma$
(\ref{Omega_int_performed}).
There are $9$ possibilities:
\be
\lbrace  w  \in [-1;\,-\xi],  &&  w  \in [-\xi;\,-\xi], \ \  w  \in [\xi;\,1] \rbrace
\nonumber \\ &&
\otimes \,
\{ v  \in (-\infty; \, -|\xi '|], \ \
v  \in [-|\xi '|; \, |\xi '|], \ \
v  \in [ |\xi '|; \, \infty) \}\,.
\label{Cases_WV}
 \ee

\begin{figure}[h]
 %\begin{center}
 \epsfig{figure=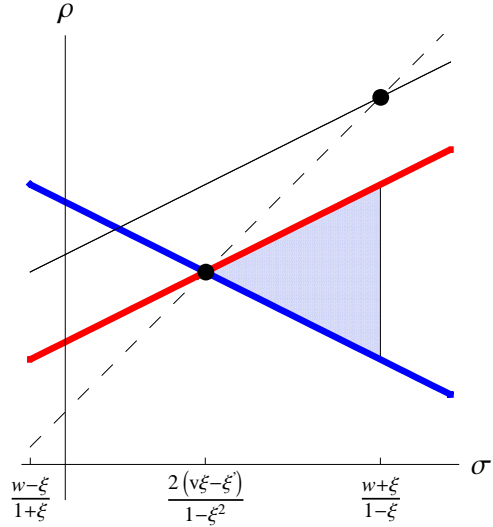 , height=7cm}
%\end{center}
  \caption{The domain of integration in $(\sigma,\,\rho)$ plane in eq.~(\ref{Omega_int_performed}) for
$-1 \le w \le -\xi$ and $\xi' \le v \le 1-\xi'+\xi$ defined by the inequalities
(\ref{triangle}) and (\ref{strip}). See explanations in the text. }
\label{Pict_Domain}
\end{figure}

Let us consider in details  the case
\be
w  \in [-1;\,-\xi]\,;  \ \ \
v  \in [ \xi';\,\infty)\,.
\label{Case_1}
\ee
The integration domain in
$(\sigma, \, \rho)$
plane is defined by the intersection of a domain specified by the inequalities
(\ref{Ineq_master1}),
(\ref{Ineq_master2_refined_1}), (\ref{Ineq_master2_refined_2}), (\ref{Ineq_master3_refined}):
\be
\rho \ge -\frac{\sigma}{2}+ \frac{v-\xi'}{1-\xi}\,;
\ \ \ \
\rho
 \le \frac{\sigma}{2}+ \frac{v+\xi'}{1+\xi}\,;
\ \ \ \
\rho \le \frac{\sigma}{2 \xi}+ \frac{\xi'}{\xi}\,;
\ \ \ \
|\rho| \le 1-\frac{|\sigma|}{2}
 \label{triangle}
\ee
with the strip
\be
\frac{w-\xi}{1+\xi} \le \sigma \le \frac{w+\xi}{1-\xi}\,.
\label{strip}
\ee
The  domain defined by the inequalities
(\ref{triangle})
and
(\ref{strip})
is presented on
Fig.~\ref{Pict_Domain}.
By the thick solid lines we show the borders of the domain
defined by the first two inequalities
(\ref{triangle}).
The thin solid line
is the border of the domain defined by the inequality
$\rho \le 1-\frac{\sigma}{2}$.
The dashed line is the border of the domain defined by the inequality
(\ref{Ineq_master3_refined}):
$\rho \le \frac{\sigma}{2 \xi}+ \frac{\xi'}{\xi}$.
The shaded area corresponds
to the resulting domain of integration in
(\ref{Omega_int_performed}) for
$-1 \le w \le -\xi$
and
$\xi' \le v \le 1-\xi'+\xi$.

The abscissa of the apex  of this  triangular domain
is
\be
\sigma = \frac{2(v  \xi -\xi')}{1-\xi^2}\,.
\ee
One may check that for
$v =\xi '$  the abscissa of the apex
coincides with the left boundary of the strip
(\ref{strip}):
\be
\left.\frac{2(v  \xi -\xi ')}{1-\xi^2} \right|_{v =\xi'}=\frac{w -\xi}{1+\xi}\,,
\ee
while for $v =1-\xi '+\xi$ it coincides with the right boundary of the strip (\ref{strip}):
\be
\left.\frac{2(v  \xi -\xi ')}{1-\xi^2} \right|_{v =1-\xi '+\xi}=\frac{w +\xi}{1-\xi}\,.
\ee
For $v  > 1-\xi '+\xi$ the apex of the triangular domain lies on the right of the strip
(\ref{strip}) and
hence has empty intersection with it.
This makes the double integral
(\ref{Omega_int_performed})
vanish for
$v  \ge 1-\xi '+\xi$
and  ensures the desired
support property of
$H (w ,v ,\xi)$.

The third inequality in
(\ref{triangle})
does not further restrict the domain since
the apex of the triangular domain belongs to the line
$\rho = \frac{\sigma}{2 \xi}+ \frac{\xi'}{\xi}$
and the triangular domain lies to the right of this line for
$0 \le \xi \le 1$.
The two inequalities
(\ref{Ineq_master1})
also do not impose additional restriction for the domain.
Indeed one may check that the
$\rho = \frac{\sigma}{2 \xi}+ \frac{\xi'}{\xi}$
intersects with $\rho=1- \frac{|\sigma|}{2}$
at $\sigma= \frac{w+\xi}{1-\xi}$.

The eight remaining cases
(\ref{Cases_WV})
can be considered according to this pattern in a completely
analogous way.
One may check that the quadruple integral
(\ref{Spectral_represent_Hi})
for
$H(w,v,\xi)$
for
$\xi \ge 0$
reduces to the following expressions:
\begin{itemize}
\item For
$w$
and
$v$
outside the domain
$  w \in [-1;\,1]$ and $v \in [-1+|\xi-\xi'| ;\,1-|\xi-\xi'|]$
the integral vanishes.
\item
For
$w  \in [-1;\,-\xi]$
and
$v \in [\xi ' ;\,  1-\xi '+\xi]$ (DGLAP-like type I domain):
\be
H(w ,\,v ,\xi)= \frac{1}{\xi^2 }
\int_{\frac{2(v  \xi -\xi ')}{1-\xi^2}}^{\frac{w +\xi}{1-\xi}} d \sigma  \int_{-\frac{\sigma }{2}+ \frac{v -\xi '}{1-\xi}}^{ \frac{\sigma }{2}+ \frac{v +\xi '}{1+\xi}}
d \rho  \, F(\sigma ,\, \rho ,\, \frac{w -\sigma }{\xi}, \, \frac{v -\rho }{\xi})\,.
\label{DGLAP_type_I_1}
\ee

\item For
$w  \in [-1;\,-\xi]$
and
$v \in [-\xi' ;\,   \xi']$ (DGLAP-like type II domain):
\be
H(w ,\,v ,\xi)= \frac{1}{\xi^2}
\int_{ \frac{w- \xi }{1+\xi }}^{\frac{w +\xi}{1-\xi}} d \sigma
\int_{-\frac{\sigma}{2}+ \frac{v -\xi'}{1+\xi}}^{ \frac{\sigma }{2}+ \frac{v +\xi '}{1+\xi}}
d \rho  \, F(\sigma ,\, \rho ,\, \frac{w -\sigma }{\xi}, \, \frac{v -\rho }{\xi})\,.
\label{DGLAP_type_II_1}
\ee

\item For
$w  \in [-1;\,-\xi]$
and
$v \in [-1+\xi'-\xi  ;\,   -\xi']$ (DGLAP-like type I domain):
\be
H(w,\,v,\xi)= \frac{1}{\xi^2}
\int_{-\frac{2(v \xi +\xi')}{1-\xi^2}}^{\frac{w+\xi}{1-\xi}} d \sigma \int_{-\frac{\sigma}{2}+ \frac{v-\xi'}{1+\xi}}^{ \frac{\sigma}{2}+ \frac{v+\xi'}{1-\xi}}
d \rho \, F(\sigma,\, \rho,\, \frac{w-\sigma}{\xi}, \, \frac{v-\rho}{\xi})\,.
\label{DGLAP_type_I_2}
\ee

\item For
$w \in [-\xi;\,\xi]$
and
$v \in [\xi' ;\, 1-\xi +\xi']$ (DGLAP-like type II domain):
\be
H(w ,\,v ,\xi)= \frac{1}{\xi^2 }
\int_{\frac{2(v  \xi -\xi ')}{1-\xi^2}}^{\frac{w +\xi}{1+\xi}} d \sigma  \int_{-\frac{\sigma }{2}+ \frac{v -\xi '}{1-\xi}}^{ \frac{\sigma }{2}+ \frac{v +\xi '}{1+\xi}}
d \rho  \, F(\sigma ,\, \rho ,\, \frac{w -\sigma }{\xi}, \, \frac{v -\rho }{\xi})\,.
\label{DGLAP_type_II_2}
\ee

\item
$w \in [-\xi;\,\xi]$
and
$v \in [-\xi' ;\, \xi']$ (ERBL-like domain):
\be
H(w ,\,v ,\xi)= \frac{1}{\xi^2}
\int_{ \frac{w- \xi }{1+\xi }}^{\frac{w +\xi}{1+\xi}} d \sigma
\int_{-\frac{\sigma}{2}+ \frac{v -\xi'}{1+\xi}}^{ \frac{\sigma }{2}+ \frac{v +\xi '}{1+\xi}}
d \rho  \, F(\sigma ,\, \rho ,\, \frac{w -\sigma }{\xi}, \, \frac{v -\rho }{\xi})\,.
\ee

\item

$w \in [-\xi;\,\xi]$
and
$v \in [-1+\xi-\xi' ;\, -\xi' ] $ (DGLAP-like type II domain):
\be
H(w ,\,v ,\xi)= \frac{1}{\xi^2}
\int_{  -\frac{2(v \xi + \xi')}{1-\xi^2}}^{\frac{w +\xi}{1+\xi}} d \sigma
\int_{-\frac{\sigma}{2}+ \frac{v -\xi'}{1+\xi}}^{ \frac{\sigma }{2}+ \frac{v +\xi '}{1-\xi}}
d \rho  \, F(\sigma ,\, \rho ,\, \frac{w -\sigma }{\xi}, \, \frac{v -\rho }{\xi})\,.
\label{DGLAP_type_II_3}
\ee

\item $w \in [\xi;\,1]$ and $v \in [-\xi';\,1-\xi+\xi']$:
the result coincides with
(\ref{DGLAP_type_II_2})
as it certainly should be since this is the part of the same
DGLAP type II domain. Note that this makes
$H(w,\,v,\,\xi)$
a smooth function for
$w=\xi$
as it should be since this line ($w_i=\xi\,\Leftrightarrow \, x_i=2 \xi$)
does not correspond to any change of evolution properties of
$H(w,\,v,\,\xi)$.

\item $w \in [\xi;\,1]$ and $v \in [\xi';-\xi']$ (DGLAP-like type I domain):
\be
H(w ,\,v ,\xi)= \frac{1}{\xi^2}
\int_{ \frac{w- \xi }{1-\xi }}^{\frac{w +\xi}{1+\xi}} d \sigma
\int_{-\frac{\sigma}{2}+ \frac{v -\xi'}{1-\xi}}^{ \frac{\sigma }{2}+ \frac{v +\xi '}{1-\xi}}
d \rho  \, F(\sigma ,\, \rho ,\, \frac{w -\sigma }{\xi}, \, \frac{v -\rho }{\xi})\,.
\label{DGLAP_type_I_3}
\ee

\item $w \in [\xi;\,1]$ and $v \in [-1+\xi-\xi' ;\, \xi']$:
the result again coincides with
(\ref{DGLAP_type_II_3})
since this is the part of the same DGLAP-like type II domain.

\end{itemize}

\section{Radyushkin type Ansatz for $\pi N$ TDAs}

In this section we discuss what could be a possible approach for
modelling of quadruple distributions
$F(\sigma, \, \rho, \, \omega,\,\nu)$
occurring in the spectral representation
(\ref{Spectral_represent_Hi}).

Employing the analogy with the case of
usual GPDs
one may assume that the profile of
$F(\sigma, \, \rho, \, \omega,\,\nu)$
in $(\sigma,\,\rho)$
space is determined by the shape of the function
$f(\sigma,\,\rho)$
to which
$\pi N$ TDA
is reduced in the limit
$\xi \rightarrow 0$.
For the moment we put aside the complicated and interesting problem of the rigorous
physical meaning of this limit. It will be discussed elsewhere.
Thus, we suggest to employ the following factorized Ansatz for quadruple
distributions:
\be
F(\sigma, \, \rho, \, \omega,\,\nu)=f(\sigma,\,\rho) \, h(\sigma,\,\rho,\,\omega,\,\nu)\,,
\label{Factorized_Ansatz}
\ee
where
$h(\sigma,\,\rho,\,\omega,\,\nu)$
is a profile function normalized according to:
\be
\int_{-1+|\sigma|}^{1-|\rho-\frac{\sigma}{2}|-|\rho+\frac{\sigma}{2}|} d \omega
\int_{-1+|\rho- \frac{\sigma}{2}|+\frac{1+\omega}{2}}^{1-|\rho+ \frac{\sigma}{2}|-\frac{1+\omega}{2}} d \nu\,
h(\sigma,\,\rho,\,\omega,\,\nu)=1\,.
\label{normalization_h}
\ee

A possible model is to
exploit further the analogy with the standard Radyushkin Ansatz for the double distributions
\cite{RDDA4}
and to assume that the
$(\omega,\,\nu)$
profile of
$h(\sigma,\,\rho,\,\omega,\,\nu)$
is determined by the shape of the asymptotic form of
the nucleon distribution amplitude:
\be
 \Phi^{\rm as}(y_1,\,y_2,\,y_3)  = \frac{15}{4} \, y_1  y_2 y_3\,.
\label{piN_DA}
\ee
The DA
(\ref{piN_DA})
is defined for $y_{1,\,2,\,3} \in [0;\,2]$ such that
$y_1+y_2+y_3=2$.

In terms of quark-diquark variables
$ \tilde{\omega}=1-y_1-y_2$ and
$\tilde{\nu}=\frac{y_1-y_2}{2}$
$\Phi^{\rm as}$
reads:
\be
\Phi^{\rm as}(\tilde{\omega},\,\tilde{\nu})=\frac{15}{4} (\tilde{\omega}+1)(\tilde{\nu}+ \frac{1-\tilde{\omega}}{2})
  (-\tilde{\nu}+ \frac{1-\tilde{\omega}}{2})\,.
\ee
Note that
\be
\int_0^2 dy_1   dy_2   dy_3 \, \delta(2-y_1-y_2-y_3) \Phi^{\rm as}(y_1,\,y_2,\,y_3) \equiv
\int_{-1}^1 d \tilde{\omega} \int_{-\frac{1-\tilde{\omega}}{2}}^{\frac{1-\tilde{\omega}}{2}} d \tilde{\nu}
\Phi^{\rm as}(\tilde{\omega},\,\tilde{\nu})=1
\ee

$\Phi^{\rm as}(\tilde{\omega},\,\tilde{\nu})$ is defined for
\be
-1 \le \tilde{\omega} \le 1\,;  \ \ \
\text{and } \ \ \
-\frac{1-\tilde{\omega}}{2} \le \tilde{\nu} \le \frac{1-\tilde{\omega}}{2}\,,
\ee
while
$h(\sigma,\,\rho,\,\omega,\,\nu)$
is defined for
\be
&&
-1+|\sigma| \le \omega \le  1-|\rho-\frac{\sigma}{2}|-|\rho+\frac{\sigma}{2}|\,; \nonumber \\ &&
 -\frac{1 - \omega}{2}+ |\rho- \frac{\sigma}{2}| \le  \nu \le \frac{1 - \omega}{2}- |\rho+ \frac{\sigma}{2}|\,.
\ee
Thus it makes sense to employ the following substitution of the variables:
\be
&&
\tilde{\omega}= \frac{\omega +\frac{1}{2} \left(\left|\rho -\frac{\sigma }{2}\right|+\left|\rho +\frac{\sigma }{2}\right|-|\sigma |\right)}{1-\frac{1}{2} \left(\left|\rho
   -\frac{\sigma }{2}\right|+\left|\rho +\frac{\sigma }{2}\right|+|\sigma |\right)}\,;
\nonumber \\ &&
  \tilde{\nu}= \frac{(1-\tilde{\omega})}{2} %\left(1-\frac{\omega }{1-|\sigma |}\right)
\frac{ 2 \nu -\left|\rho -\frac{\sigma }{2}\right|+\left|\rho +\frac{\sigma }{2}\right|}{1-\omega-\left|\rho -\frac{\sigma }{2}\right|-\left|\rho +\frac{\sigma }{2}\right|}\,.
\label{substitution_profile}
\ee
This results in the following expression for the profile function
$h(\sigma,\,\rho,\, \omega,\, \nu)$:
\be
%&&
h(\sigma,\,\rho,\, \omega,\, \nu) %\nonumber \\ &&
= \frac{15}{16} \frac{ \left(1+2 \nu -\omega -2 \left|\rho -\frac{\sigma }{2}\right|\right) \left(1-2 \nu -\omega -2 \left|\rho +\frac{\sigma
   }{2}\right| \right) (1 -|\sigma |+\omega)}{  \left(1-\frac{1}{2} \left(\left|\rho -\frac{\sigma }{2}\right|+\left|\rho
   +\frac{\sigma }{2}\right|+|\sigma |\right) \right)^5}\,.
%\nonumber \\ &&
\label{Profile_h}
\ee
One may check that the profile function
(\ref{Profile_h})
satisfies the normalization condition
(\ref{normalization_h}).
It is extremely interesting to note that
in terms of the initial spectral parameters
$\alpha_1$, $\alpha_2$, $\alpha_3$
and
$\beta_1$, $\beta_2$, $\beta_3$
satisfying
$\alpha_1+\alpha_2+\alpha_3=-1$
and
$\beta_1+\beta_2+\beta_3=0$
the profile function (\ref{Profile_h})
can be rewritten in the very symmetric form:
\be
&&
\left.h(\beta_1, \beta_2, \beta_3\,; \alpha_1, \alpha_2, \alpha_3)
\right|_{\sum_i \beta_i=0 \atop \sum_i \alpha_i=-1}
=\frac{15}{4}
\frac{ \prod_{i=1}^3 (1+\alpha_i-|\beta_i|)}{(1-\frac{1}{2} (|\beta_1|+|\beta_2|+|\beta_3|))^5}
\Big|_{\sum_i \beta_i=0 \atop \sum_i \alpha_i=-1}
\,.
\nonumber \\ &&
\ee

The inverse transformation
(\ref{substitution_profile})
reads
\be
&&
\omega=
 \tilde{\omega} \left(1- \frac{1}{2} \left( \left|\rho -\frac{\sigma }{2}\right|+\left|\rho +\frac{\sigma }{2}\right|+|\sigma | \right)\right)-\frac{1}{2} \left(\left|\rho -\frac{\sigma
   }{2}\right|+\left|\rho +\frac{\sigma }{2}\right|-|\sigma |\right)\,;
\nonumber \\ &&
\nu=
\tilde{\nu} \left(1- \frac{1}{2} \left( \left|\rho -\frac{\sigma }{2}\right|+\left|\rho +\frac{\sigma }{2}\right|+|\sigma | \right)\right)+
\frac{1}{2} \left(\left|\rho -\frac{\sigma }{2}\right|-\left|\rho +\frac{\sigma }{2}\right|\right)\,.
\ee
This allows to easily compute the integrals occurring in the calculation of
$(N-n,\,n)$-th Mellin moments
($N \ge n \ge 0$)
in
$(w,\,v)$ of $\pi N$ TDAs:
\be
\int_{-1+|\sigma|}^{1-|\rho-\frac{\sigma}{2}|-|\rho+\frac{\sigma}{2}|} d \omega
\int_{-1+|\rho- \frac{\sigma}{2}|+\frac{1+\omega}{2}}^{1-|\rho+ \frac{\sigma}{2}|-\frac{1+\omega}{2}} d \nu\,
\omega^{N-n} \nu^n
h(\sigma,\,\rho,\,\omega,\,\nu)\,.
\ee

In principle one may also think of a more intricate profile function.
In fact any particular function
$\Phi(\tilde{\omega},\tilde{\nu})$
normalized according to
\be
\int_{-1}^1 d \tilde{\omega} \int_{-\frac{1-\tilde{\omega}}{2}}^{\frac{1-\tilde{\omega}}{2}}
d \tilde{\nu} \Phi(\tilde{\omega},\tilde{\nu})=1
\ee
will   define some profile function
$h(\sigma,\,\rho,\, \omega,\, \nu)$
after the substitution
(\ref{substitution_profile})%
\footnote{
However, one has to make certain assumptions on the endpoint behavior of
the function
$f(\sigma,\,\rho)$
to which
$\pi N$ TDA
is reduced in the limit
$\xi \rightarrow 0$.}.
{\it E.g.} taking $\Phi(\tilde{\omega},\tilde{\nu}) \sim (\tilde{\omega}+1)^{b_1}(\tilde{\nu}+ \frac{1-\tilde{\omega}}{2})^{b_2}
  (-\tilde{\nu}+ \frac{1-\tilde{\omega}}{2})^{b_3}$
would lead to the natural generalization of the
$b$
parameter dependent Radyushkin profile familiar for usual GPDs.

It is interesting  also to consider the most simple possible profile with no distortion
in $(\omega, \nu)$ directions:
\be
\Phi(\tilde{\omega},\tilde{\nu})  = \delta(\tilde{\omega}) \delta(\tilde{\nu}).
\label{Delta_profile}
\ee
Contrary to the case of usual GPDs for which the counterpart of the
profile (\ref{Delta_profile}) leads to $\xi$-independent Ansatz the
resulting $\pi N$ TDA preserves the minimal necessary $\xi$ dependence.
Indeed
$\tilde{\omega}=0$
and
$\tilde{\nu}=0$
does not imply
$\omega=0$
and
$\nu=0$
and hence the $\xi$-dependence introduced through two $\delta$-functions
in
(\ref{Spectral_represent_Hi})
is preserved and generates the proper $\xi$-dependent domain of definition for
the resulting $\pi N$ TDA
(\ref{The_whole_domain}).
Unfortunately the model with the profile
(\ref{Delta_profile})
turns out to be pathological since it leads to
$\pi N$
TDAs which
are not continuous at the cross-over lines
$v= \pm \xi'$
and
$w=-\xi$
separating ERBL-like and DGLAP-like type I, II domains. This makes
impossible the calculation of the amplitude of the hard exclusive process in question given
 by convolution of
$\pi N$
TDA with the appropriate hard part
(see \cite{Lansberg:2007ec}).
Indeed the imaginary part of the corresponding amplitude is given
by the values of  $\pi N$ TDA
at the cross-over lines
$v= \pm \xi'$
and
$w=-\xi$.

For the moment as a toy model we are going to employ the factorized Ansatz
(\ref{Factorized_Ansatz})
with the profile function
(\ref{Profile_h}). It is a good point now to discuss a possible model
for the function
$f(\sigma, \rho)$
that  is the second ingredient of the factorized Ansatz
(\ref{Factorized_Ansatz}).
In the  limit $\xi \rightarrow 0$
$\pi N$ TDA  reduces to this function:
\be
H(w,v,\xi=0)= f(w,v)\,.
\ee

%%%%%%%%%%%%%%%%%%%%%%%%%%%%%%%%%%%%%%%%%%%%%%%%%%%%%%%%%%%%%%%%%%%%%%
The requirements of convergence of integrals
(\ref{DGLAP_type_I_1})--(\ref{DGLAP_type_I_3})
for
$\pi N$ TDA
impose some restriction on the
behavior of the function
$f(\sigma, \rho)$
on the border of its domain of definition.
It turns out that
$f(\sigma, \rho)$
should vanish at least as a certain power of the relevant variables
at the borders of its domain of definition.
%%%%%%%%%%%%%%%%%%%%%%%%%%%%%%%%%%%%%%%%%%%%%%%%%%%%%%%%%%%%%%%%%%%%%%
Thus for the function
$f(\sigma,\,\rho)$
we suggest the following simple form:
\be
&&
f(\sigma,\,\rho) = \theta(-1 \le \sigma \le 1) \,
\theta(-1+  \frac{|\sigma|}{2}  \le \rho \le 1-  \frac{|\sigma|}{2} ) \,\nonumber \\ &&
\times
\frac{40}{47} \left(1-\sigma ^2\right) \left((\rho -1)^2-\frac{\sigma
   ^2}{4}\right) \left((\rho +1)^2-\frac{\sigma ^2}{4}\right)\,.
%\nonumber \\ &&
\label{model_for_f}
\ee
In terms of the initial spectral parameters
$\beta_i$ 
satisfying
$\sum_i\beta_i=0$
(\ref{model_for_f})
can be rewritten as:
\be
\left.f(\beta_1,\,\beta_2,\,\beta_3)\right|_{\sum_i\beta_i=0}=
  \frac{40}{47} \, \prod_{i=1}^3 \theta(|\beta_i|\le 1) (1-\beta_i^2) \big|_{\sum_i\beta_i=0}\,.
\label{model_for_f_beta}
\ee
The function
$f(\sigma,\,\rho)$
vanishes on the border of its domain of definition and
is normalized according to
\be
\int_{-1}^1 d \sigma \int_{-1+  \frac{|\sigma|}{2}  }^{1- \frac{|\sigma|}{2}  } f(\sigma,\,\rho)=1\,.
\label{Norm_f}
\ee
Let us stress that we employ the normalization
(\ref{Norm_f})
only for our toy model.
Advanced modelling of $\pi N$ TDAs aiming the quantitative description
of the physical observables would certainly require more complicated form of
$f(\sigma, \rho)$.

The normalization for
the nucleon to pion TDAs can be derived either from the soft pion limit
or from the lattice calculations of several first Mellin moments of
$\pi N$
TDAs or from the comparison with the results of
\cite{Pasquini:2009ki}.
On the other hand it can be computed considering the light baryon exchange contributions into the Mellin moments of
$\pi N$
TDAs using the phenomenological values say of
$g_{\pi NN}$
and
$g_{\pi N \Delta}$
couplings. The normalization can also in principle be established
directly form the experimental measurements of the cross-section
once the scaling behavior would be found reasonable.

\begin{figure}[H]
 \begin{center}
\epsfig{figure=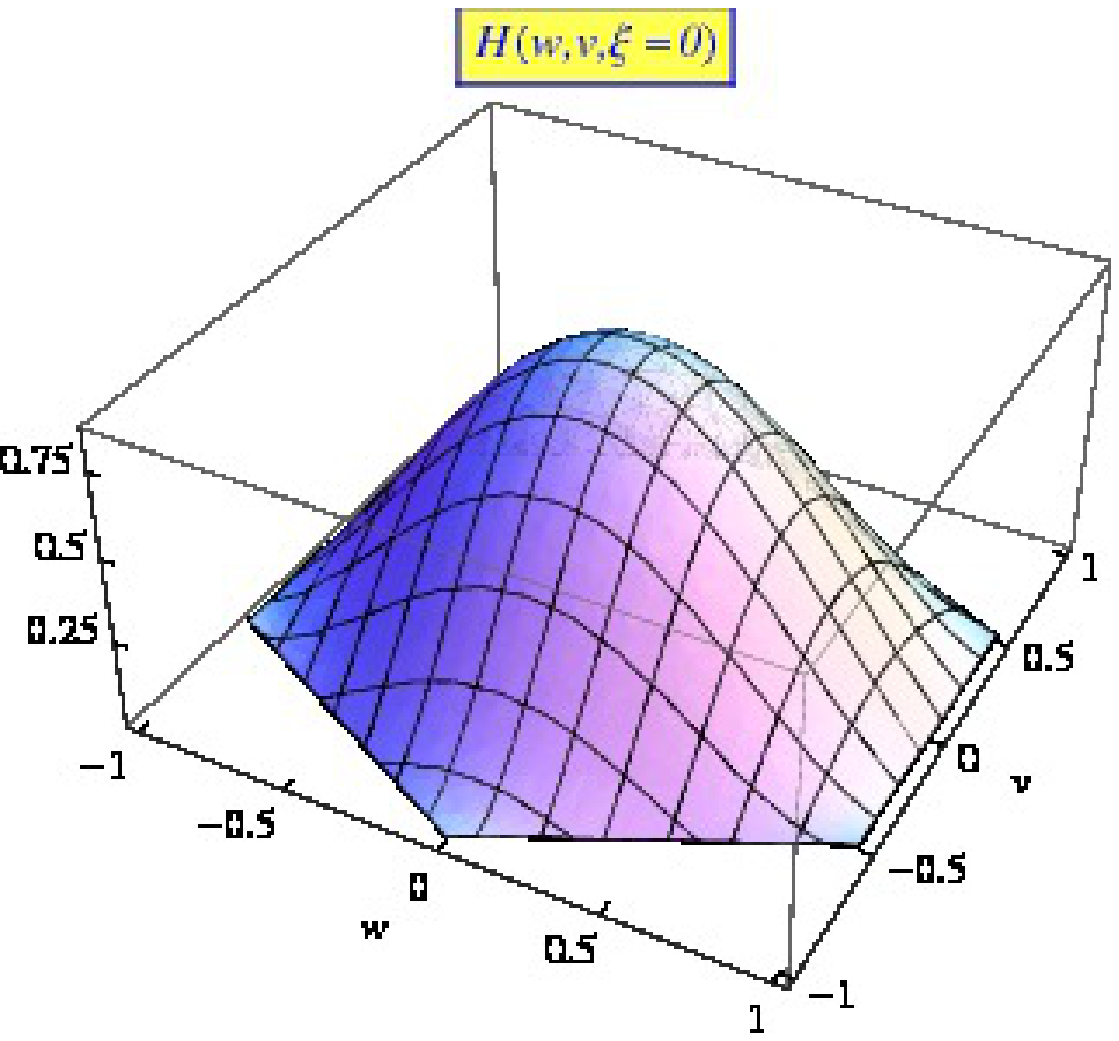, height=6cm}
\epsfig{figure=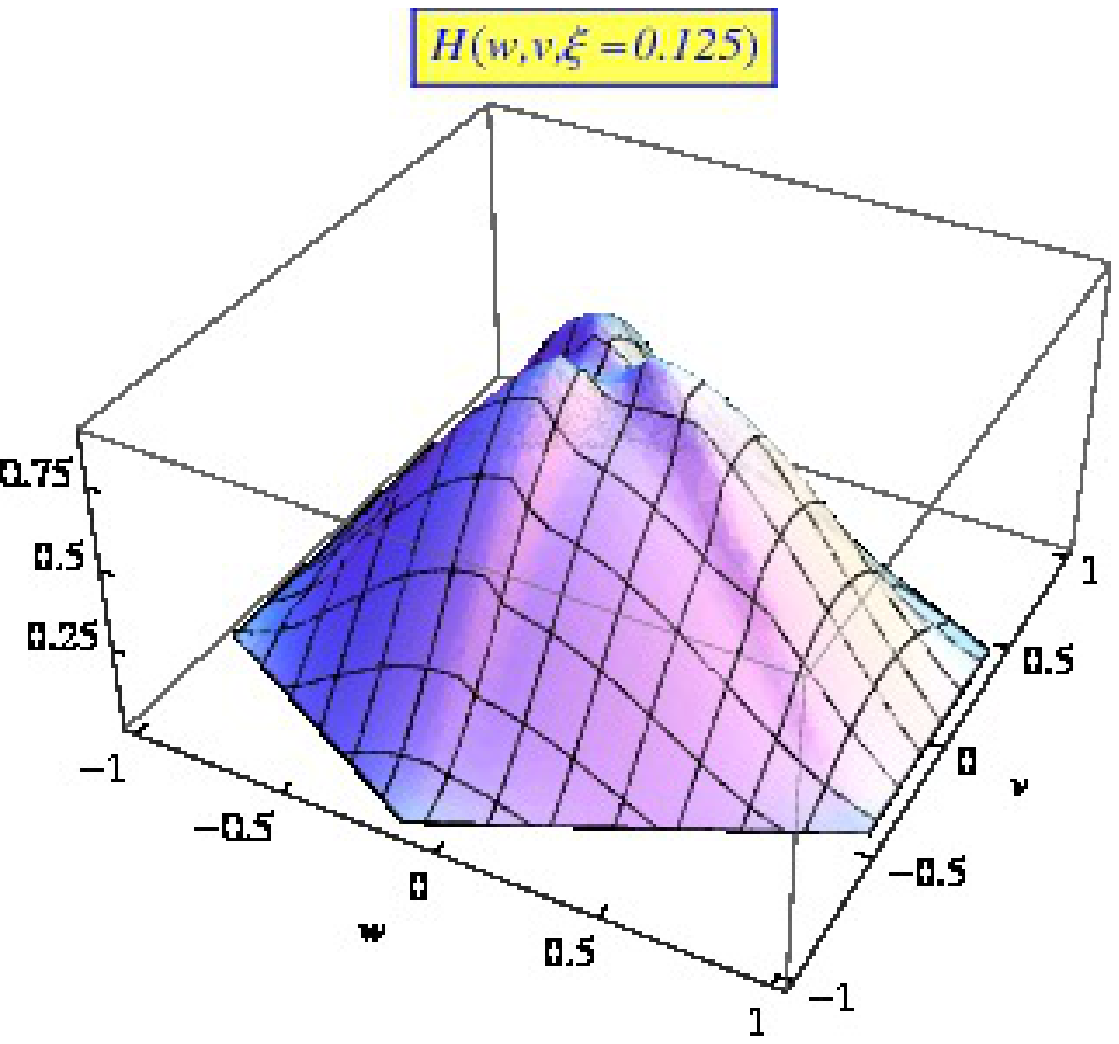, height=6cm}
\epsfig{figure=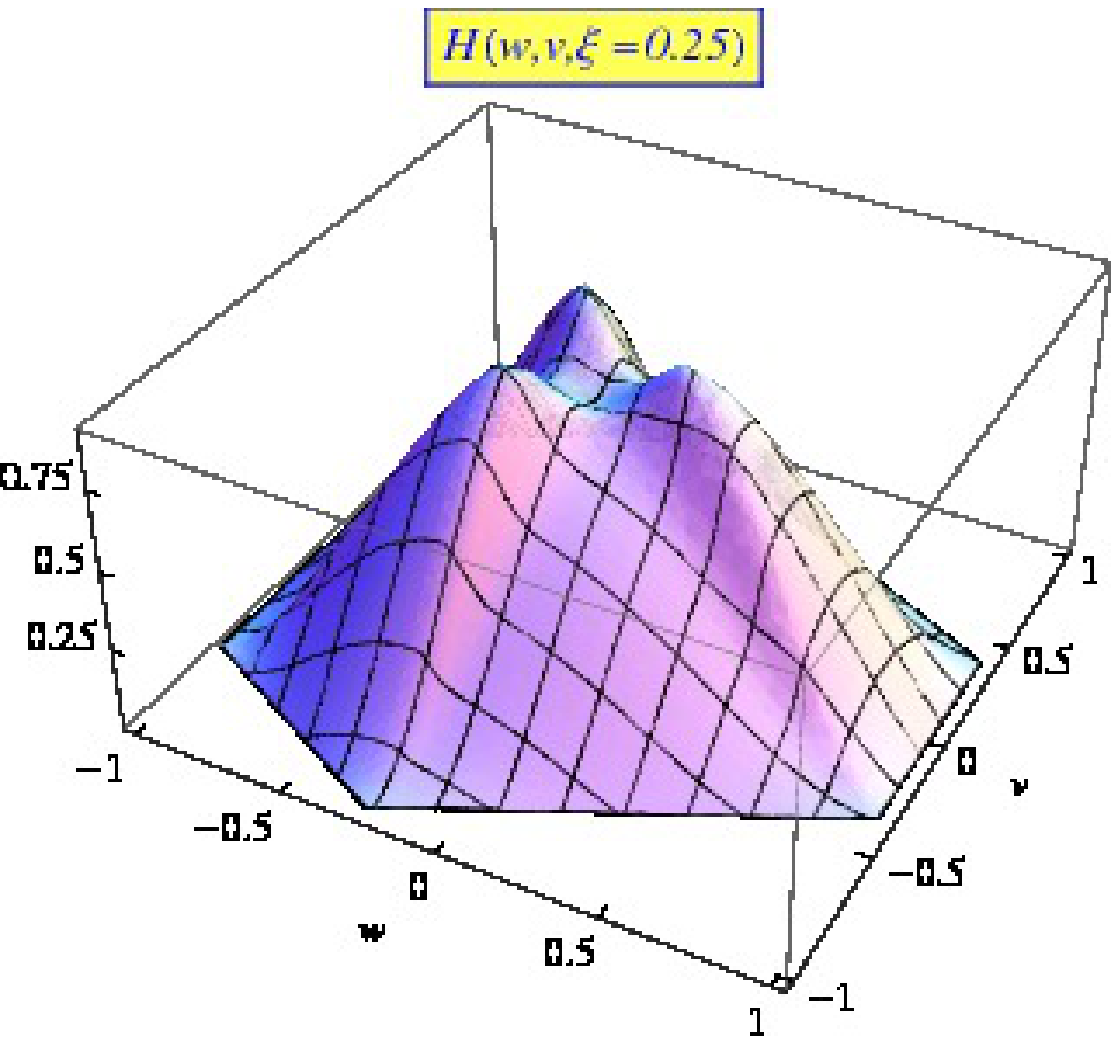, height=6cm}
\epsfig{figure=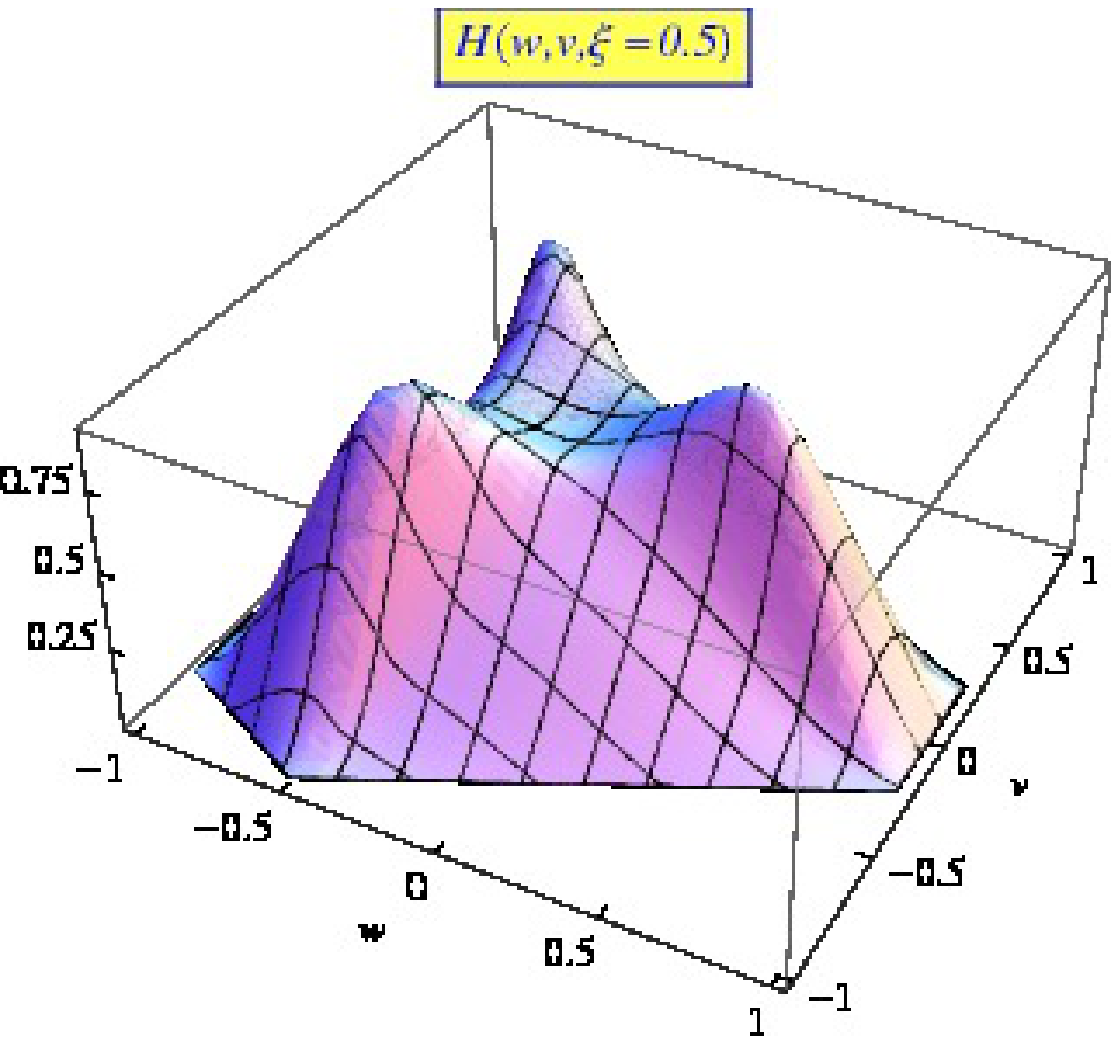, height=6cm}
\epsfig{figure=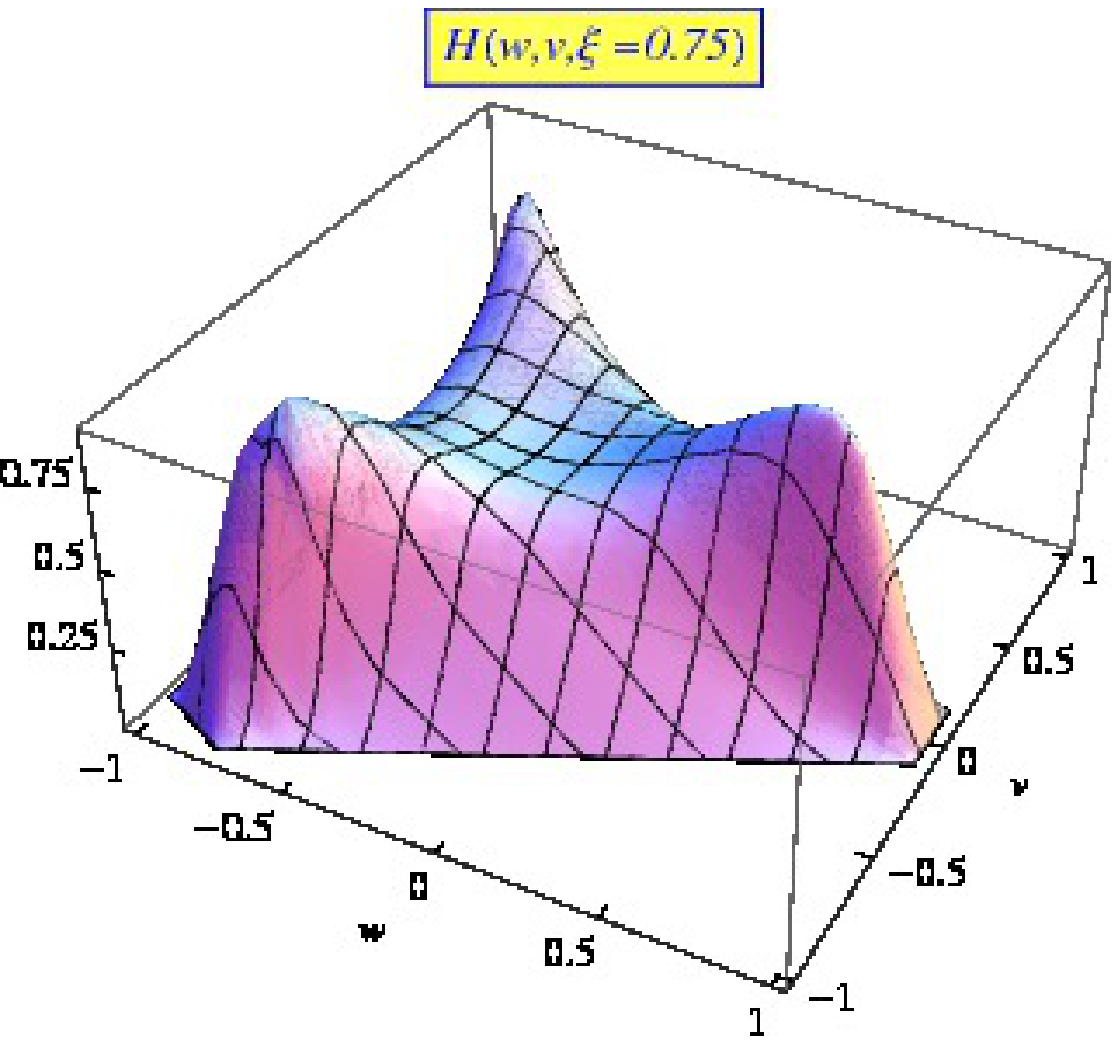, height=6cm}
\epsfig{figure=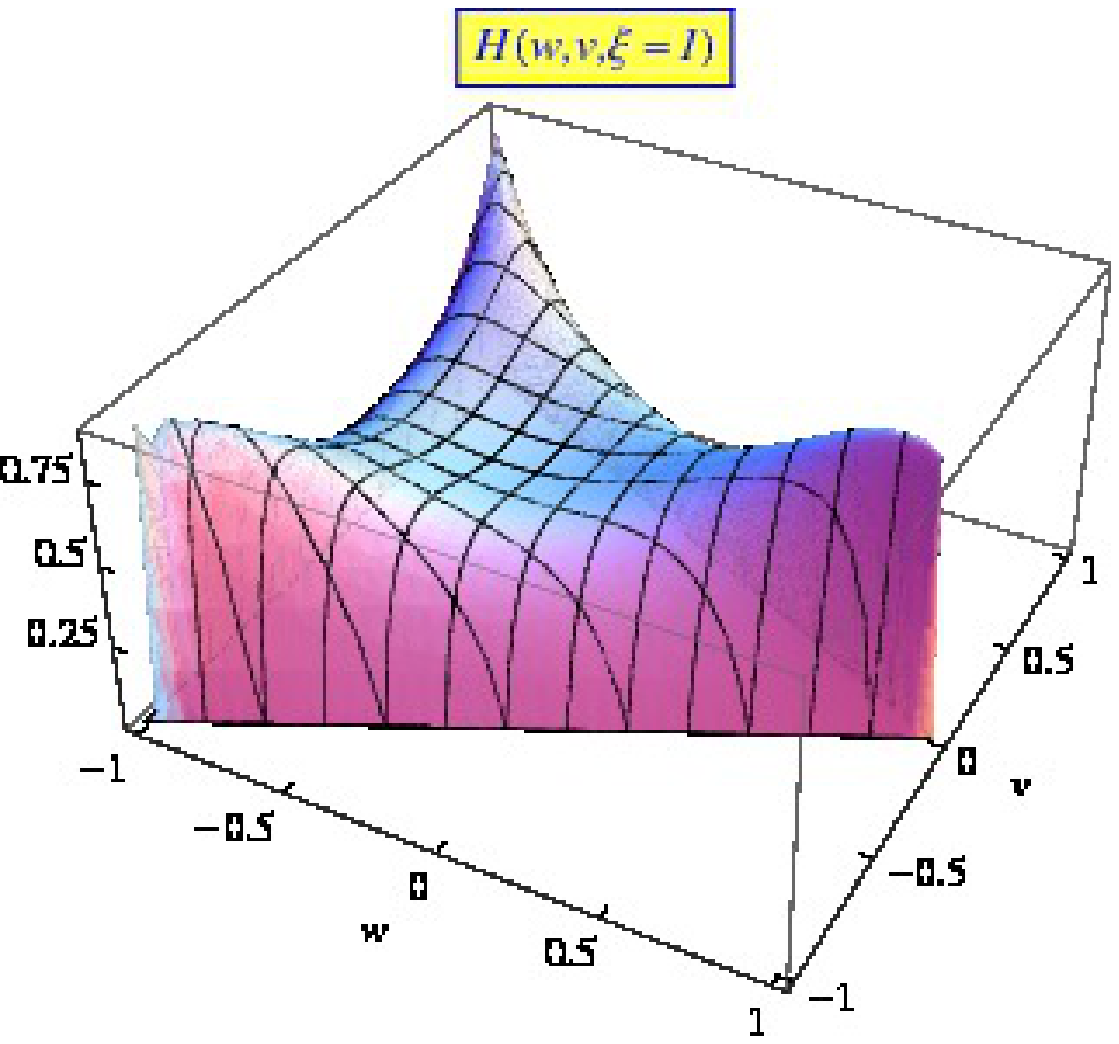, height=6cm}
  \caption{The contribution into $\pi N$ TDA
$H(w_3,v_3,\xi) \equiv H (w ,v ,\xi)$
as a function of
$w$
and
$v$
for different values of $\xi$ computed using the
factorized Ansatz (\ref{Factorized_Ansatz}) with the profile function (\ref{Profile_h}) and
$f(\sigma,\,\rho)$
given by (\ref{model_for_f}).}
\label{Fig_xi1}
\end{center}
\end{figure}

On Figure~\ref{Fig_xi1}
we show the results of the calculation of the contribution
$H(w_3,v_3,\xi) \equiv H (w ,v ,\xi)$
as a function of
$w$
and
$v$
for different values of
$\xi$
computed with the help of the factorized Ansatz
(\ref{Factorized_Ansatz})
with the profile
(\ref{Profile_h})
and
$f(\sigma,\,\rho)$
given by the toy model
(\ref{model_for_f}).

Note that for
$\xi=1$
the TDA
$H(w ,v ,\xi)$
does not vanish at the corners of its domain of definition. This is potentially dangerous
since this may lead to the break up of the factorization property of the hard exclusive process in question.
Fortunately this problem is an artefact of our oversimplified toy model
(\ref{model_for_f_beta})
for the forward limit of
$\pi N$ TDA. It was checked that taking
$f(\sigma, \rho)$
that vanishes quadratically at the borders of the domain of definition
\be
\left.f(\beta_1,\,\beta_2,\,\beta_3)\right|_{\sum_i\beta_i=0}=
\frac{4410}{3167} \, \prod_{i=1}^3 \theta(|\beta_i|\le 1) (1-\beta_i^2)^2 \big|_{\sum_i\beta_i=0}
\ee
leads to a vanishing
$\pi N$
TDA  at the corners of its domain of definition for
$\xi=1$.

\begin{figure}[H]
 \begin{center}
\epsfig{figure=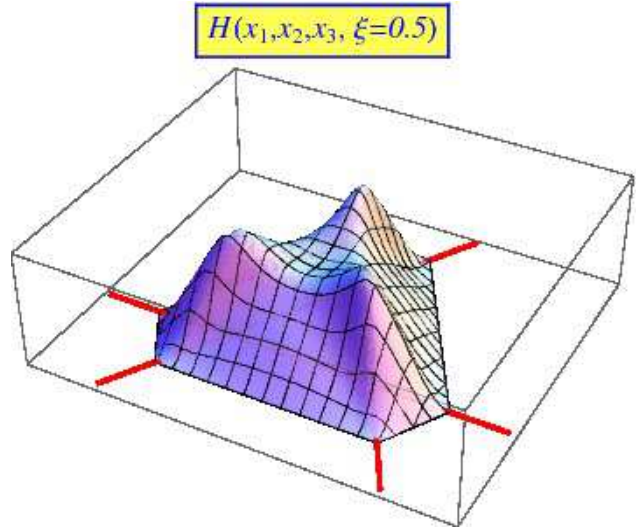, height=7cm}
  \caption{  $\pi N$ TDA
$ H(x_1 ,x_2 ,x_3,\xi)$
as a function of
$x_1$,
$x_2$ and $x_3$ ($x_1+x_2+x_3=2 \xi$)
for $\xi=0.5$ in barycentric coordinates. By thick solid lines we show the continuation of the edges of the
equilateral triangle that border the ERBL-like domain {\it cf.} Fig.~\ref{BaryCentric_Fig}.}
\label{Fig_bary}
\end{center}
\end{figure}

On Figure~\ref{Fig_bary} we show
$\pi N$
TDA
$ H(x_1 ,x_2 ,x_3,\xi)$
for $\xi=0.5$
as a function of three dependent light-cone momentum fractions
$x_1$,
$x_2$ and $x_3$ ($x_1+x_2+x_3=2 \xi$)
in the barycentric coordinates. By thick solid lines we show the continuation of the edges of the
equilateral triangle which form the ERBL-like domain {\it cf.} Fig.~\ref{BaryCentric_Fig}.

\section{Conclusions}
\label{Sec_Conclusions}

The non-perturbative part of hard processes involving hadrons is encoded in various
universal partonic distributions (parton distribution functions, fragmentation
functions, distribution amplitudes and their generalizations). Waiting for a
complete understanding of the dynamics of quark and gluon confinement in hadrons,
one should model these distributions in agreement with general
requirements of the underlying field theory such as
Lorentz invariance and causality.
Spectral representation of hadronic matrix
elements offers an elegant way to address this program.
The double distribution representation for GPDs became the basis for various
successful phenomenological GPD models.

In this paper we  introduced the notion of quadruple distributions and constructed
the spectral representation for the transition distribution amplitudes involving
three parton correlators which arise in the description of baryon to meson
%and baryon to photon
transitions.
We also generalized  Radyushkin's factorized Ansatz for the case of quadruple distributions and provided
an explicit expression for the corresponding profile function.
Analogously to the case of GPDs the shape of the corresponding profile function is supposed to be fixed by the asymptotic
form of the nucleon distribution amplitude.
Our model also requires the knowledge of nucleon to meson TDAs in the forward limit as input
quantities. Contrarily to the GPD case, the nucleon to meson TDAs suffer from the fact that there is no illuminating forward limit.
This problem requires further investigation. For a moment, we suggest to employ a simple shape of nucleon to
meson TDAs in the forward limit assuming
that they are fixed by their behavior at the borders of their domain of definition.
Our construction opens the way to quantitative modeling of baryon-meson and baryon-photon TDAs
in their complete domain of  definition.

Let us emphasize that for the moment we have not included any $D$-term like contributions to the
spectral representation of the nucleon to meson TDAs in terms of quadruple distributions.
Indeed the results of
\cite{Belitsky:2000vk} and of Chapter~3.8 of \cite{BelRad}
give us confidence that the eventual $D$-term like contributions to TDAs can be included by
means of complementing the spectral density in
(\ref{Spectral_for_GPDs_x123}) with additional terms
proportional to powers of
$\xi$.
The subsequent analysis can be performed according to the same pattern.

Let us also point  out that our method can be  generalized
for the case of $4$-quark correlators important for the description of higher twist contributions.

\section*{Acknowledgements}
%We would especially like to thank
We are   thankful to Igor~Anikin, Jean-Philippe Lansberg, Anatoly~Radyushkin and
Samuel~Wallon
for many discussions and helpful comments.
K.S. also acknowledges much the partial support by Consortium Physique des Deux Infinis (P2I).
This work was supported by the Polish Grant N202 249235.

\setcounter{section}{0}
\setcounter{equation}{0}
\renewcommand{\thesection}{\Alph{section}}
\renewcommand{\theequation}{\thesection\arabic{equation}}

\section{A useful constrained integral}
\label{App_Useful_int}

Let us consider the constrained triple integral
\be
I(a,\,b,\,c)= \int_{-a}^a d \alpha_1
\int_{-b}^b d \alpha_2
\int_{-c}^c d \alpha_3
\, \delta(\alpha_1+\alpha_2+\alpha_3+1) \, f(\alpha_1,\,\alpha_2,\,\alpha_3)\,,
\label{Master_Int}
\ee
where
 $a \ge 0\,; \, a \le 1$,  $b \ge 0\,; \, b \le 1$,  $c \ge 0\,; \, c \le 1$.
We introduce the natural coordinates
$\omega_3$
and
$\nu_3$:
\be
\alpha_1= \nu_3+ \frac{-1-\omega_3}{2}; \ \ \ \alpha_2= -\nu_3+ \frac{-1-\omega_3}{2}; \ \ \ \alpha_3=\omega_3\,.
\ee
In the  natural coordinates
$\omega_3$
and
$\nu_3$
the integration in
(\ref{Master_Int})
is over the intersection of three stripes:
\be
 \nonumber
-c \le  & \omega_3  & \le c\,; \\
-a+\frac{1+\omega_3}{2} \le & \nu_3 & \le a+\frac{1+\omega_3}{2}\,; \nonumber  \\
-b-\frac{1+\omega_3}{2} \le & \nu_3 & \le b-\frac{1+\omega_3}{2}\,.
\label{Stripes}
\ee

One may check that for $ a \ge b$ the integral (\ref{Master_Int})  can be rewritten as
\be
&&
I(a,\,b,\,c)= \int_{-1-a-b}^{-1-a+b} d \omega_3\, \theta(\omega_3+c) \, \theta(c-\omega_3) \int_{-b-\frac{1+\omega_3}{2}}^{a+\frac{1+\omega_3}{2}} d \nu_3\,
  f(\omega_3,\,\nu_3) \nonumber \\ &&
+
\int_{-1-a+b}^{-1+a-b} d \omega_3\, \theta(\omega_3+c) \, \theta(c-\omega_3)  \int_{-b-\frac{1+\omega_3}{2}}^{b-\frac{1+\omega_3}{2}} d \nu_3\,
  f(\omega_3,\,\nu_3)
\nonumber \\ &&
+
\int_{-1+a-b}^{-1+a+b} d \omega_3\, \theta(\omega_3+c) \, \theta(c-\omega_3)  \int_{-a+\frac{1+\omega_3}{2}}^{b-\frac{1+\omega_3}{2}} d \nu_3 \, f(\omega_3,\,\nu_3)\,.
\label{Integral_ab_1}
\ee
%%%%%%%%%%%%%%%%%%%%%%%%%%%%%%%%%%%%%%%%%%%%%%%%%%%%%%%%%%%%%%%%%%%%%%%%%%%%%%%%%%%%%%%%%%%%%%%%%%%%%%%%%%%%%%%%%%%%%%%%%%%%%%%%%---------->
Analogously for $ b \ge a  $ the integral (\ref{Master_Int})  can be rewritten as
\be
&&
I(a,\,b,\,c)= \int_{-1-a-b}^{-1+a-b} d \omega_3  \, \theta(\omega_3+c) \, \theta(c-\omega_3) \int_{-b-\frac{1+\omega_3}{2}}^{a+\frac{1+\omega_3}{2}} d \nu_3\,
 f(\omega_3,\,\nu_3) \nonumber \\ &&
+
\int_{-1-b+a}^{-1+b-a} d \omega_3 \, \theta(\omega_3+c) \, \theta(c-\omega_3) \int_{-a+\frac{1+\omega_3}{2}}^{a+\frac{1+\omega_3}{2}} d \nu_3\,
  f(\omega_3,\,\nu_3)
\nonumber \\ &&
+
\int_{-1+b-a}^{-1+a+b} d \omega_3\, \theta(\omega_3+c) \, \theta(c-\omega_3)  \int_{-a+\frac{1+\omega_3}{2}}^{b-\frac{1+\omega_3}{2}} d \nu_3
  \, f(\omega_3,\,\nu_3)\,.
\label{Integral_ab_2}
\ee

In order to be able to perform the integral
(\ref{Master_Int})
we
need to specify the intersection of three stripes
(\ref{Stripes}).
The results
(\ref{Integral_ab_1})
and
(\ref{Integral_ab_2})
are obtained for arbitrary positive
$a$, $b$
and
$c$.
Let us now take into the account that
\be
a=1-|\beta_1|; \ \ \ b=1-|\beta_2|; \ \ \ c=1-|\beta_3|
\ee
with
$|\beta_i| \le 1$
and
$\beta_1+\beta_2+\beta_3=0$.

\begin{itemize}
\item
Let us first consider the case when
$\beta_i$s
belong to the domain
$D_1$
(\ref{Def_3_Domains}).
In this domain we have
$|\beta_1|=|\beta_2|+|\beta_3|$
and thus $a=b+c-1$.
So in the domain
$D_1$
the following inequalities are respected:
\be
0 \le a \le 1\,, \;  \ \ \  0 \le b \le 1\,, \ \ \  0 \le a-b+1 \le 1 \,.
\ee
One may check that these inequalities result in
\be
c \ge -1+a+b\,; \ \ \ -c=-1+b-a\,; \ \ \ \text{and} \ \ \ b \ge a\,.
\label{Ineq_1}
\ee
Thus employing
(\ref{Integral_ab_2})
we get
\be
\left. I(a,b,c) \right|_{D_1}= \int_{-c}^{-1+a+b} d \omega_3
\int_{-a+\frac{1+\omega_3}{2}}^{ b-\frac{1+\omega_3}{2}} d \nu_3   \, f(\omega_3,\,\nu_3)\,.
\ee

\item Analogously, in the domain
$D_2$
we have
$|\beta_2|=|\beta_1|+|\beta_3|$
and thus $b=a+c-1$.
The inequalities
\be
0 \le a \le 1\,, \;  \ \ \  0 \le b \le 1\,, \ \ \  0 \le b-a+1 \le 1
\ee
result in
\be
c \ge -1+a+b\,; \ \ \ -c=-1+a-b\,; \ \ \ \text{and} \ \ \ a \ge b\,.
\label{Ineq_2}
\ee
Employing
(\ref{Integral_ab_1})
we get
\be
\left. I(a,b,c) \right|_{D_2}= \int_{-c}^{-1+a+b} d \omega_3
\int_{-a+\frac{1+\omega_3}{2}}^{ b-\frac{1+\omega_3}{2}} d \nu_3   \, f(\omega_3,\,\nu_3)\,.
\ee

\item Finally, let us consider the case when
$\beta_i$
belong to the domain
$D_3$.
In this domain we have
$|\beta_3|=|\beta_1|+ |\beta_2|$
and hence
$c=a+b-1$.
Thus in the domain
$D_3$
the following inequalities are respected:
\be
0 \le a \le 1\,,   \ \ \ 0 \le b \le 1\,, \ \ \   0 \le a+b-1 \le 1 \,.
\ee
One may check that in this domain
\be
-c \ge -1+b-a\,; \ \ \ -c \ge -1+a-b\,; \ \ \ c=-1+a+b\,.
\label{Ineq_3}
\ee
Thus
independently of
$a  \ge b$
or
$a  \le b$
the integral over the intersection of three stripes
(\ref{Integral_ab_1}) or (\ref{Integral_ab_2})
is again reduced to
\be
\left. I(a,b,c) \right|_{D_3}= \int_{-c}^{-1+a+b} d \omega_3
\int_{-a+\frac{1+\omega_3}{2}}^{ b-\frac{1+\omega_3}{2}} d \nu_3   \, f(\omega_3,\,\nu_3)\,.
\ee
\end{itemize}

\section{Case $\xi < 0$ }
\label{App_xi_le_0}
For completeness in this Appendix we present the result for
$\pi N$
TDA
$H(w,v,\xi)$
in the ERBL-like and DGLAP-like type I and II domains
for the case
$-1 \le \xi <0$
which is useful {\it e.g.} for
$\bar{N}N \rightarrow \pi \gamma^*$ in the
forward region
\cite{LPS}.

\begin{itemize}
\item For
$w$
and
$v$
outside the domain
$-1 \le w \le 1$ and $-1+|\xi-\xi'| \le v \le 1-|\xi-\xi'|$
the integral vanishes.

\item
For
$w  \in [-1;\,\xi]$
and
$v \in [\xi ' ;\,  1-\xi '+\xi]$ (DGLAP-like type II domain):
\be
H(w ,\,v ,\xi)= \frac{1}{\xi^2 }
\int^{\frac{2(v  \xi -\xi ')}{1-\xi^2}}_{\frac{w +\xi}{1-\xi}} d \sigma  \int^{-\frac{\sigma }{2}+ \frac{v -\xi '}{1-\xi}}_{ \frac{\sigma }{2}+ \frac{v +\xi '}{1+\xi}}
d \rho  \, F(\sigma ,\, \rho ,\, \frac{w -\sigma }{\xi}, \, \frac{v -\rho }{\xi})\,.
\label{DGLAP_type_II_1_xi_le_0}
\ee
%Ok

\item For
$w  \in [-1;\,\xi]$
and
$v \in [-\xi' ;\,   \xi']$ (DGLAP-like type I domain):
\be
H(w ,\,v ,\xi)= \frac{1}{\xi^2}
\int_{ \frac{w+\xi }{1-\xi }}^{\frac{w -\xi}{1+\xi}} d \sigma
\int^{-\frac{\sigma}{2}+ \frac{v -\xi'}{1+\xi}}_{ \frac{\sigma }{2}+ \frac{v +\xi '}{1+\xi}}
d \rho  \, F(\sigma ,\, \rho ,\, \frac{w -\sigma }{\xi}, \, \frac{v -\rho }{\xi})\,.
\label{DGLAP_type_I_1_xi_le_0}
\ee
%Ok

\item For
$w  \in [-1;\,\xi]$
and
$v \in [-1+\xi'-\xi  ;\,   -\xi']$ (DGLAP-like type II domain):
\be
H(w,\,v,\xi)= \frac{1}{\xi^2}
\int^{-\frac{2(v \xi +\xi')}{1-\xi^2}}_{\frac{w+\xi}{1-\xi}} d \sigma \int^{-\frac{\sigma}{2}+ \frac{v-\xi'}{1+\xi}}_{ \frac{\sigma}{2}+ \frac{v+\xi'}{1-\xi}}
d \rho \, F(\sigma,\, \rho,\, \frac{w-\sigma}{\xi}, \, \frac{v-\rho}{\xi})\,.
\label{DGLAP_type_II_2_xi_le_0}
\ee
%Ok

\item For
$w \in [\xi;\,-\xi]$
and
$v \in [-\xi' ;\, 1-\xi' +\xi]$ (DGLAP-like type II domain)
the result coincides with (\ref{DGLAP_type_II_1_xi_le_0}).
%\be
%H_i(w ,\,v ,\xi)= \frac{1}{\xi^2 }
%\int^{\frac{2(v  \xi -\xi ')}{1-\xi^2}}_{\frac{w +\xi}{1-\xi}} d \sigma  \int^{-\frac{\sigma }{2}+ \frac{v -\xi '}{1-\xi}}_{ \frac{\sigma }{2}+ %\frac{v +\xi '}{1+\xi}}
%d \rho  \, F_i(\sigma ,\, \rho ,\, \frac{w -\sigma }{\xi}, \, \frac{v -\rho }{\xi})
%\label{DGLAP_type_II_2_xi_le_0}
%\ee
%Ok

\item
$w \in [\xi;\,-\xi]$
and
$v \in [\xi' ;\, -\xi']  $ (ERBL-like domain):
\be
H(w ,\,v ,\xi)= \frac{1}{\xi^2}
\int_{ \frac{w+ \xi }{1-\xi }}^{\frac{w -\xi}{1-\xi}} d \sigma
\int^{-\frac{\sigma}{2}+ \frac{v -\xi'}{1-\xi}}_{ \frac{\sigma }{2}+ \frac{v +\xi '}{1-\xi}}
d \rho  \, F(\sigma ,\, \rho ,\, \frac{w -\sigma }{\xi}, \, \frac{v -\rho }{\xi})\,.
\ee
%Ok

\item
$w \in [\xi;\,-\xi]$
and
$v \in [-1+\xi'-\xi ; \,  \xi']  $ (DGLAP-like type II domain):
the result coincides with
(\ref{DGLAP_type_II_2_xi_le_0}).
%\be
%H_i(w ,\,v ,\xi)= \frac{1}{\xi^2}
%\int_{  -\frac{2(v \xi + \xi')}{1-\xi^2}}^{\frac{w +\xi}{1+\xi}} d \sigma
%\int_{-\frac{\sigma}{2}+ \frac{v -\xi'}{1+\xi}}^{ \frac{\sigma }{2}+ \frac{v +\xi '}{1-\xi}}
%
%d \rho  \, F_i(\sigma ,\, \rho ,\, \frac{w -\sigma }{\xi}, \, \frac{v -\rho }{\xi})
%\label{DGLAP_type_II_3}
%\ee

%%%%%%%%%%%%%%%%%%%%%%%%%%%%%%%%%%%%%%%%%%%%%%%%%%%%%%%%%%%%%%%%%%%%%%%%%%%%%%%%%%%%%%%%%%%%%%%%%%%

\item $w \in [-\xi;\,1]$ and $v \in [-\xi';\,1-\xi+\xi']$ (DGLAP-like type I domain):
\be
H(w ,\,v ,\xi)= \frac{1}{\xi^2}
\int_{ \frac{w+ \xi }{1+\xi }}^{\frac{2(v \xi -\xi')}{1-\xi^2}} d \sigma
\int^{-\frac{\sigma}{2}+ \frac{v -\xi'}{1-\xi}}_{ \frac{\sigma }{2}+ \frac{v +\xi '}{1+\xi}}
d \rho  \, F(\sigma ,\, \rho ,\, \frac{w -\sigma }{\xi}, \, \frac{v -\rho }{\xi})\,.
\ee

\item $w \in [-\xi;\,1]$ and $v \in [\xi';-\xi']$ (DGLAP-like type II domain):
\be
H(w ,\,v ,\xi)= \frac{1}{\xi^2}
\int_{ \frac{w+ \xi }{1+\xi }}^{-\frac{2(v \xi +\xi')}{1-\xi^2}} d \sigma
\int^{-\frac{\sigma}{2}+ \frac{v -\xi'}{1-\xi}}_{ \frac{\sigma }{2}+ \frac{v +\xi '}{1-\xi}}
d \rho  \, F(\sigma ,\, \rho ,\, \frac{w -\sigma }{\xi}, \, \frac{v -\rho }{\xi})\,.
\ee

\item $w \in [\xi;\,1]$ and $v \in [-1+\xi-\xi' ; \,   \xi']$ (DGLAP-like type I domain):
 \be
H(w ,\,v ,\xi)= \frac{1}{\xi^2}
\int_{ \frac{w+ \xi }{1+\xi }}^{\frac{w -\xi}{1-\xi}} d \sigma
\int^{-\frac{\sigma}{2}+ \frac{v -\xi'}{1+\xi}}_{ \frac{\sigma }{2}+ \frac{v +\xi '}{1-\xi}}
d \rho  \, F(\sigma ,\, \rho ,\, \frac{w -\sigma }{\xi}, \, \frac{v -\rho }{\xi})\,.
\ee

\end{itemize}

\end{document}